\documentclass[aps,prd, reprint,nofootinbib]{revtex4-2}
\pdfoutput=1
\usepackage{amsmath,
amssymb,
amsthm,
amsxtra,
babel,
braket,
booktabs,
graphicx,
dsfont,
slashed,
multirow,
siunitx,
stackengine,
xcolor,
xparse,
}

\usepackage[hidelinks]{hyperref}

\usepackage{makecell}
\usepackage[utf8]{inputenc}

\newcommand{\eqb}{\begin{equation}}
\newcommand{\eqe}{\end{equation}}
\newcommand{\dmb}{\begin{displaymath}}
\newcommand{\dme}{\end{displaymath}}

\newcommand{\eab}{\begin{eqnarray}}
\newcommand{\eae}{\end{eqnarray}}

\newcommand{\be}{\begin{equation}}
\newcommand{\ee}{\end{equation}}

\graphicspath{{Notebooks/}}

\newcolumntype{L}{>{$}c<{$}}

\RenewDocumentCommand\be{}{\begin{equation}}
\RenewDocumentCommand\ee{}{\end{equation}}
\NewDocumentCommand\der{}{\mathrm{d}}

\NewDocumentCommand\lsim{}{\lesssim}

\NewDocumentCommand\intkern{}{\int\kern-5pt}

\NewDocumentCommand\lagr{}{\mathcal{L}}
\NewDocumentCommand\abgd{}{\alpha\beta\gamma\delta}

\NewDocumentCommand\plushc{}{+\mathrm{H.c.}}
\NewDocumentCommand\ope{}{\mathcal{O}}
\NewDocumentCommand\caes{}{^{133}\text{Cs}}
\NewDocumentCommand\carb{}{^{12}\text{C}}

\stackMath

\begin{document}

\title{Searching for new physics from SMEFT and leptoquarks at the P2 experiment}

\author{Ingolf Bischer}
\affiliation{Max-Planck-Institut f\"ur Kernphysik, Saupfercheckweg 1,  69117 Heidelberg,  Germany}
\author{P. S. Bhupal Dev}
\affiliation{Department of Physics and McDonnell Center for the Space Sciences,  Washington University, St.\ Louis, MO 63130, USA}
\author{Werner Rodejohann}
\affiliation{Max-Planck-Institut f\"ur Kernphysik, Saupfercheckweg 1,  69117 Heidelberg,  Germany}
\author{Xun-Jie Xu}
\affiliation{Institute of High Energy Physics, Chinese Academy of Sciences, Beijing 100049, China}
\author{Yongchao Zhang}
\affiliation{School of Physics, Southeast University, Nanjing 211189, China}

\begin{abstract}
The P2 experiment aims at high-precision measurements of the parity-violating asymmetry in elastic electron-proton and
electron-$^{12}$C scatterings with longitudinally polarized electrons.
We discuss here the sensitivity of P2 to leptoquarks, which within the P2 energy range can be described in the language of Standard Model Effective Field Theory (SMEFT). We give the expected P2 limits on the SMEFT operators and on the leptoquark parameters, which will test energy scales up to 15 TeV. In many cases those limits exceed current constraints from LHC and atomic parity violation (APV) experiments. We also demonstrate that degeneracies of different SMEFT operators can partially be resolved by use of APV experiments and different targets (protons and $^{12}$C) at P2. Moreover, we show that P2 could  confirm or resolve potential  tensions between the theoretical and experimental determinations of the weak charge of $^{133}$Cs. 
\end{abstract}

\maketitle

\section{\label{sec:intro}Introduction}
Leptoquarks are colored scalar or vector particles that couple to the 
quarks and leptons 
of the Standard Model (SM), see Ref.~\cite{Dorsner:2016wpm} for a review. They arise in many theories beyond the SM, including Grand Unified Theories~\cite{Pati:1974yy, Senjanovic:1982ex, Murayama:1991ah, Frampton:1991ay, Hewett:1997ce, Dorsner:2005fq, Assad:2017iib, Murgui:2021bdy}, string and M-theories~\cite{Gershtein:1999gp}, $R$-parity-violating supersymmetry~\cite{Barbier:2004ez} and radiative neutrino mass models~\cite{AristizabalSierra:2007nf, Babu:2010vp, Dorsner:2017wwn,Cai:2017jrq, Babu:2019mfe}.
They are frequently used candidates to explain the current anomalies in the muon anomalous magnetic moment and in semileptonic $B$-decays~\cite{Deshpande:2012rr,Sakaki:2013bfa,Alonso:2015sja, Calibbi:2015kma, Freytsis:2015qca, Bauer:2015knc, Fajfer:2015ycq,Li:2016vvp,Becirevic:2016yqi,Sahoo:2016pet,Hiller:2016kry,Bhattacharya:2016mcc, Popov:2016fzr,Barbieri:2016las,ColuccioLeskow:2016dox,Crivellin:2017zlb,Hiller:2017bzc,Cai:2017wry,Altmannshofer:2017poe,Buttazzo:2017ixm, DiLuzio:2017vat,Calibbi:2017qbu,Blanke:2018sro,Becirevic:2018afm,Crivellin:2018yvo,Heeck:2018ntp,Angelescu:2018tyl,Iguro:2018vqb,Chauhan:2018lnq,Fornal:2018dqn,Aydemir:2019ynb,Cornella:2019hct,Popov:2019tyc,Bigaran:2019bqv,DaRold:2019fiw,Crivellin:2019dwb,Bigaran:2020jil,Altmannshofer:2020axr, Dev:2020zcy,Crivellin:2020tsz,Babu:2020hun,Angelescu:2021lln, Nomura:2021oeu,Lee:2021jdr,Du:2021zkq, Ban:2021tos,Dev:2021ipu, Bigaran:2021kmn, Belanger:2021smw}; see Ref.~\cite{Fischer:2021sqw} for a review. In general, the quantum numbers of leptoquarks are governed by the quantum numbers of SM particles  involved~\cite{Buchmuller:1986zs}. As the SM is a chiral theory, hence violates parity, leptoquarks can be expected to influence measurable parity violation. More generally, parity-violating interactions at low-energy scales from new physics at large mass scales can be described by operators of SM Effective Field Theory (SMEFT)~\cite{Buchmuller:1985jz,Grzadkowski:2010es}.

In this paper we focus on the P2 experiment, which will measure the parity-asymmetry in elastic electron-proton or electron-$^{12}$C scatterings 
at the upcoming Mainz 
Energy-recovering Superconducting Accelerator (MESA) facility~\cite{Becker:2018ggl}.  The crucial observable is the parity violating cross section asymmetry $(\sigma_L - \sigma_R)/(\sigma_L + \sigma_R)$ for the scattering of  longitudinally polarized electrons off the targets. The expectation is that the SM parity asymmetry can be measured with 1.4\% (0.3\%) relative uncertainty for a proton ($^{12}$C) target, which can be used to constrain beyond SM contributions~\cite{Becker:2018ggl,Dev:2021otb}. Our goal in this paper is to use these expected values to set prospective limits on SMEFT coefficients and leptoquark masses and couplings to first-generation SM fermions.

As the momentum exchange relevant for P2 is less than 100 MeV~\cite{Becker:2018ggl} which is  very  small compared to the allowed mass of leptoquarks, one can approximate their interactions in terms of effective SMEFT operators.
Hence, we start by obtaining the prospective limits on relevant SMEFT operators in \autoref{tab:bounds} and \autoref{tab:bounds-smeft}, which can be translated into constraints on specific models inducing these operators, as shown in  \autoref{tab:lq-bounds}. Subsequently, we use the identification with SMEFT operators to map out expected bounds on leptoquark parameters.  
We confront those leptoquark bounds with current constraints, most notably from atomic parity violation (APV) using $^{133}$Cs~\cite{Sahoo:2021thl} and from leptoquark production at the LHC in the pair-production~\cite{ATLAS:2020dsk, Diaz:2017lit}, Drell-Yan~\cite{ATLAS:2020yat} and single resonant production~\cite{Crivellin:2021egp} channels. As we will show in Figures~\ref{fig:lqbounds-scalar} and \ref{fig:lqbounds-vector}, in many cases the P2 limits will supersede current constraints, if the mass of the leptoquark exceeds roughly 2 TeV.  We also confirm that certain SMEFT operators lead to an indistinguishable effect in a given observable. In this case one can partially resolve this degeneracy by taking advantage of the fact that different observables use different amounts of up- and down-quarks in the target material, as illustrated in \autoref{fig:smeftbounds}. We therefore stress the complementarity of using protons and $\carb$ at P2, as well as $\caes$ in APV to better disentangle up- and down-quark interactions.  

We note that there have been a number of studies on low-energy
parity violating effects of new physics~\cite{Davoudiasl:2012ag,Davoudiasl:2012qa,Erler:2014fqa,  Davoudiasl:2014kua,  Dzuba:2017puc,  Safronova:2017xyt,  Hao:2020zlz, Sahoo:2020dcv} 
and in particular on the P2 sensitivity to new physics~\cite{Dev:2021otb,Crivellin:2021bkd}.
While Ref.~\cite{Dev:2021otb} focuses on low-mass gauge bosons
and Ref.~\cite{Crivellin:2021bkd} considers broadly a variety of constraints
from low-energy observables to collider searches, this work presents
a dedicated study on the P2 sensitivity to new physics that can be
handled via the SMEFT, with an application to the leptoquark scenario. Our main new observation is the importance
of using $^{12}{\rm C}$ target to break the degeneracy between
up- and down-quark couplings.

The rest of the paper is organized as follows. In \autoref{sec:origin} we lay out the formalism of SMEFT operators and leptoquarks with focus on parity violation in the first generation. In \autoref{sec:constr} we map the fundamental quark-level couplings to nucleon and nucleus couplings, calculate the parity asymmetry and set limits on SMEFT energy scales  and leptoquark masses. \autoref{sec:other} compares those limits to current ones from APV and LHC. The advantage of using different targets is illustrated in \autoref{sec:deg} by taking two effective couplings at a time. 
Our conclusions are presented in \autoref{sec:concl}. Mapping of the SMEFT coefficients to the flavor basis is done in \autoref{sec:app-flavorbasis}, and the parametrization of the form factors is presented in \autoref{sec:app-formfactors}.

\section{High-energy origins of parity violation in electron-hadron scattering}\label{sec:origin}

New physics at large mass scales may lead to deviations from the expected parity violation at scattering experiments. In this section, we describe two frequently investigated scenarios of new physics. The first is the model-independent approach of parametrizing new physics as effective operators in the SMEFT. The second approach is considering minimal scenarios of leptoquarks as explicit particle extensions of the SM.
\subsection{Standard Model Effective Field Theory}

If one considers new physics at some high mass scale $\Lambda\gg m_W$, a suitable framework to encode different possible effects at energies well below that scale is given by effective field theories. For the SM it is convenient to apply SMEFT~\cite{Brivio:2017vri}. Besides the SM Lagrangian, this effective theory consists of a series of non-renormalizable operators of higher mass-dimension $d\geq5$, such that we can write
\be\label{eq:eftexpansion}
\lagr_{\mathrm{eff}} = \lagr_{\mathrm{SM}}+\sum_i\sum_{n\geq5}\frac{1}{\Lambda^{n-4}}C_i\mathcal{O}^{(n)}_i\,, 
\ee
where $C_i$ denotes the dimensionless Wilson coefficient of the operator $\mathcal{O}_i$ defined e.g.\ in Refs.~\cite{Buchmuller:1985jz,Grzadkowski:2010es}. Operators of dimension $4+n$ are suppressed by respective factors of $\Lambda^{-n}$. This expansion is supposed to be applicable at interaction energies $q^2\ll \Lambda^2$, while new particles associated with $\Lambda$ may only be produced on-shell at much 
higher energy scales.

The SMEFT operators of lowest dimension giving rise to $eeuu$ or $eedd$ interactions relevant at the P2 experiment are, in the usual terminology~\cite{Grzadkowski:2010es}, the parity-violating operators\footnote{Other works on the parity-violating operators can be found e.g.\ in Refs.~\cite{Falkowski:2017pss,Boughezal:2021kla}.} $\ope_{lq(1)}$, $\ope_{lq(3)}$, $\ope_{lu}$, $\ope_{ld}$, $\ope_{qe}$, $\ope_{eu}$, $\ope_{ed}$, and the parity-conserving operators $\ope_{ledq}$, $\ope_{lequ(1)}$, and $\ope_{lequ(3)}$. Starting with the parity-violating operators the effective Lagrangian can be written as
\begin{eqnarray}
\label{eq:eff-lagrangian}
\lagr_\text{PV}&=& \frac{1}{\Lambda^2}\sum_{f=u,d}\sum_{X=L,R}(\overline{e}\gamma_\mu P_X e) \nonumber \\
&& \left[C_{XVf}(\overline{f}\gamma^\mu f)+C_{XAf}(\overline{f}\gamma^\mu\gamma^5 f)\right],
\end{eqnarray}
where $e$, $u$, and $d$ denote electron, up quark and down quark mass eigenstates. The chirality projectors are given by $P_{L,R} = \frac 12 (1 \mp \gamma_5)$. The coefficients $C_{XYf}$ are related to the coefficients of SMEFT operators in the mass basis via 
\begin{subequations}
\begin{alignat}{2}
\label{eq:EFT-matching-nucleon}
C_{LVu}&= \frac12(C_{lq(1)}-C_{lq(3)} + C_{lu})\,,\\
C_{LVd}&= \frac12(C_{lq(1)}+C_{lq(3)}+ C_{ld})\,,\\
C_{LAu}&= \frac12(-C_{lq(1)}+C_{lq(3)} + C_{lu})\,,\\
C_{LAd}&= \frac12(-C_{lq(1)}-C_{lq(3)}+ C_{ld})\,,\\
C_{RVu}&= \frac12(C_{qe}+C_{eu})\,,\\
C_{RVd}&= \frac12(C_{qe}+C_{ed})\,,\\
C_{RAu}&= \frac12(-C_{qe}+C_{eu})\,,\\
C_{RAd}&= \frac12(-C_{qe}+C_{ed})\,,
\end{alignat}
\end{subequations}
where first-generation indices $ee11$ are implied. For a mapping to the flavor basis, see \autoref{sec:app-flavorbasis}.
The parity-conserving scalar and tensor interactions read
\be
\label{eq:eff-lagrangian-spt}
\begin{split}
\lagr_\text{PC}&=\frac{1}{\Lambda^2}\sum_{f=u,d}(\overline{e}P_L e)\left[C_{Sf}(\overline{f} f)+C_{Pf}(\overline{f}\gamma^5 f)\right]\\
&\quad
+(\overline{e}\sigma_{\mu\nu}P_L e)\left[C_{Tf}\overline{f}\sigma^{\mu\nu}P_L f\right] \plushc,
\end{split}
\ee
with the mass-basis coefficients
\begin{subequations}
\begin{alignat}{2}
\label{eq:EFT-matching-nucleon-sp}
C_{Su} &= C_{Pu} = -\frac12 C_{lequ(1)}\,,\\
C_{Sd} &= -C_{Pd} = \frac12 C_{ledq}\,,\\
C_{Tu} &=  - C_{lequ(3)}\,, \\
C_{Td} &= 0\,.
\end{alignat}
\end{subequations}
We have checked that these definitions are consistent with previous investigations of parity violation from SMEFT~\cite{Boughezal:2021kla}. 
Since (pseudo)scalar and tensor interactions are parity-conserving, the P2 sensitivity of these interactions is poor compared to other probes. For this reason we focus on the parity-violating vector and axial vector interactions.

\subsection{Leptoquarks}

Leptoquarks are hypothetical scalar or vector particles which generically contribute to parity-violating interactions. In this work we follow the naming convention of Ref.\  \cite{Buchmuller:1986zs} and consider all leptoquarks which can give rise to effective quark-electron interactions. These are listed, along with their quantum numbers, in \autoref{tab:leptoquarks}. For convenience, we write all of them as fundamental representations of SU(3)$_\text{C}$.

The contributions to the parity-violating SMEFT operators from integrating out heavy leptoquarks read
\begin{subequations}
\begin{alignat}{2}
\label{eq:lq-couplings-first}
C_{lq(1)} &= -\frac{1}{4}|s_{1L}|^2-\frac{3}{4}|s_3|^2
+ \frac12 |u_{1L}|^2 +\frac32|u_3|^2 \,,
\\
C_{lq(3)} &= +\frac{1}{4}|s_{1L}|^2-\frac{1}{4}|s_3|^2
+ \frac12 |u_{1L}|^2 -\frac12|u_3|^2 \,,
\\
C_{eu} &= -\frac{1}{2}|s_{1R}|^2
+ |\widetilde{u}_1|^2\,,
\\
C_{ed} &= -\frac{1}{2}|\widetilde{s}_{1}|^2
+ |u_{1R}|^2\,,
\\
\label{eq:lq-coupling-cqe}
C_{qe} &= \frac{1}{2}|r_{2R}|^2
- |v_{2R}|^2\,,
\\
C_{lu} &= \frac{1}{2}|r_{2L}|^2
- |\widetilde{v}_{2}|^2\,,
\\
C_{ld} &= \frac{1}{2}|\widetilde{r}_{2}|^2
- |v_{2L}|^2\,,
\label{eq:lq-couplings-last}
\end{alignat}
\end{subequations}
where we have identified $\Lambda=m_\text{LQ}$ (the same for all leptoquarks). 
The couplings are defined through the interaction Lagrangians
\begin{subequations}
\begin{alignat}{2}
\begin{split}
\label{eq:theory-leptoquarks1}
\lagr_{F=2} &=
\left(s_{1L}\,\overline{q}^a\epsilon^{ab} (l^c)^b + s_{1R}\,\overline{u_R}e_R^c 
\right)S_1 \\
&\quad + \widetilde{s}_1\,\overline{d_R}e^c\, \widetilde{S}_1 
+s_{3}\overline{q}^a(\vec\tau)^{ab} \epsilon^{bd}(l^c)^d\,\vec{S}_3
\\
&\quad +\left(v_{2R}\,\overline{q}^a\gamma_\mu e_R^c + v_{2L}
\,\overline{d_R}\gamma_\mu (l^c)^a\right)V_2^{\mu,a} \\
&\quad +
\widetilde{v}_{2}
\,\overline{u_R}\gamma_\mu (l^c)^a
\widetilde{V}_2^{\mu,a}
\plushc,
\end{split}
\\
\begin{split}\label{eq:theory-leptoquarks2}
\lagr_{F=0} &=
\left(r_{2R}\,\overline{q}^b e_R + r_{2L}\,\overline{u_R}\,l^a\epsilon^{ab} \right)R_2^b \\
&\quad +\left(\widetilde{r}_{2}\,\overline{d_R}\,l^a\epsilon^{ab} 
\right){R_2^b}'\\
&\quad +\left(u_{1L}\,\overline{q}\gamma_\mu l + u_{1R}\,\overline{d_R}\gamma_\mu e_R \right)U_1^\mu \\
&\quad +\widetilde{u}_1\,\overline{u_R}\gamma_\mu e_R\, {U_1^\mu}'
+u_{3}\,\overline{q}\,\vec\tau \gamma_\mu\,l\,\vec{U}_3^\mu
\plushc,
\end{split}
\end{alignat}
\end{subequations}
where $a,b$ denote SU(2)$_L$ indices,  $\epsilon^{ab}$ is the Levi-Civita symbol, and $\vec\tau=(\tau_1,\tau_2,\tau_3)$ are the Pauli matrices.
Generally, we refer to any single mass by $m_\text{LQ}$ and any single coupling by $g_\text{LQ}$.

\begin{table}
{\centering
\caption{Leptoquarks along with their quantum numbers investigated in this work. We use the convention where electric charge $Q=I_3+Y$ with weak isospin component $I_3$ and hypercharge $Y$; $F=3B+L$ denotes fermion number, with baryon number $B$ and lepton number $L$.}
\label{tab:leptoquarks}
}
\begin{tabular}{LLLLLL}
\toprule
 & F=3B+L &\text{Spin} & \mathrm{SU(3)}_\mathrm{C} & \mathrm{SU(2)}_L & \mathrm{U(1)}_Y\\
\midrule
S_1 	&2 &0 &3 & 1 &-1/3 \\
\widetilde{S}_1  &2 &0 &3 & 1 &-4/3\\
S_3  &2 &0 &3 & 3 &-1/3\\
V_2  &2 &1 &3 & 2 &-5/6 \\
\widetilde{V}_2 &2 &1 &3 & 2 & 1/6 \\
R_2  &0 &0 &3 & 2 &7/6 \\
\widetilde{R}_2 &0 &0 &3 & 2 &1/6 \\
U_1  &0 &1 &3 & 1 &2/3 \\
\widetilde{U}_1  &0 &1 &3 & 1 &5/3\\
U_3  &0 &1 &3 & 3 &2/3 \\
\bottomrule
\end{tabular}
\end{table}

Collider searches have ruled out leptoquarks of masses below about \SI{1.8}{TeV}, as discussed in \autoref{sec:collider}. Therefore, we can focus on a regime where the full leptoquark propagators can be approximated using $|q^2|\ll m_\text{LQ}^2$:  
\begin{subequations}
 \begin{alignat}{2}\label{eq:G-UV}
G_{\mu\nu}^{\text{vector}}(q) &= i\frac{-g_{\mu\nu} + \frac{q_\mu q_\nu}{m_\text{LQ}^2}}{q^2-m_\text{LQ}^2}\approx i g_{\mu\nu}\frac{1}{m_\text{LQ}^2}\,, \\
G^{\text{scalar}}(q) &= i \frac{1}{q^2-m_\text{LQ}^2}\approx -i \frac{1}{m_\text{LQ}^2} \,.
\end{alignat}
\end{subequations}
 Therefore, for large leptoquark masses P2 is sensitive to the combination $g_\text{LQ}/m_\text{LQ}$.

\section{Setting constraints}\label{sec:constr}

\subsection{Mapping fundamental couplings to nucleon and nucleus couplings}

In order to connect the fundamental couplings with quarks to those with protons or, more generally, with nuclei, we need to apply the nuclear matrix elements and introduce form factors. Following the usual convention explained e.g.\ in Ref.~\cite{DelNobile:2021icc}, we write for the vector and axial-vector currents 
\begin{subequations}
\label{eq:formfactors}
\begin{alignat}{2}
\begin{split}
&\braket{N(p')|\overline{f}\gamma^\mu f |N(p)} \\
&= 
\overline{u}(p')\left[F_1^{f,N}(q^2)\gamma^\mu +
F_2^{f,N}(q^2)\frac{i\sigma^{\mu\nu}q_\nu}{2m_N}\right]u(p)\,,
\end{split}\\
\begin{split}
& \braket{N(p')|\overline{f}\gamma^\mu\gamma^5 f |N(p)} \\
&= 
\overline{u}(p')\left[G_A^{f,N}(q^2)\gamma^\mu\gamma^5 
+G_P^{f,N}(q^2)\frac{\gamma^5q^\mu}{2m_N}
\right]u(p)\,,
\end{split}
\end{alignat}
\end{subequations}
where $q=p_e-k_e$ is the difference of initial and final state electron momenta, $f=u,d,s$ (note that we include $s$ quarks) and $N=p,n$ for nucleons. 
Assuming isospin symmetry, the form factors related to the vector matrix element can be related to the electromagnetic Dirac and Pauli form factors $F_1^{N}$ and $F_2^{N}$ as
\begin{subequations}
\label{eq:istdasso?}
\begin{alignat}{2}
F_i^{u,p}&=F_i^{d,n} = 2F_i^p + F_i^n + F_i^{s,p}\,,\\
F_i^{d,p}&=F_i^{u,n} = F_i^p + 2F_i^n + F_i^{s,p}\,.
\end{alignat}
\end{subequations}
The form factors $F_i^N$ can be rewritten in terms of the electric and magnetic Sachs form factors $G_E^N$ and $G_M^N$ as
\begin{subequations}
\label{eq:def-sachs}
\begin{alignat}{2}
F_1^N(q^2)&=\frac{G_E^N(q^2) - \frac{q^2}{4m_N^2}G_M^N(q^2)}{
1-q^2/4m_N^2}\,,\\
F_2^N(q^2)&=\frac{G_M^N(q^2) - G_E^N(q^2)}{
1-q^2/4m_N^2}\,.
\end{alignat}
\end{subequations}
The same holds true for the strange form factors $F_i^{s,p}$ which can be expressed in terms of Sachs form factors $G_E^s$ and $G_M^s$ analogous to Eq.~\eqref{eq:def-sachs}. We use the same parametrization of the form factors as Ref.~\cite{Becker:2018ggl}. Details are given in \autoref{sec:app-formfactors}. 
Turning to the axial form factors, we take~\cite{DelNobile:2021icc}
\begin{subequations}
\begin{alignat}{2}
& G_A^{q,N}=\Delta_q^{(N)}\,,\\
& G_P^{q,N}=-4m_N^2\left(\frac{a_{q,\pi}^N}{q^2-m_\pi^2}
+ \frac{a_{q,\eta}^N}{q^2-m_\eta^2}\right), 
\end{alignat}
\end{subequations}
where
\begin{eqnarray}
a_{u,\pi}^p &=& -a_{d,\pi}^p = \frac12g_A^3 \,, \nonumber \\
a_{s,\pi}^N &=& 0 \,, \nonumber \\
a_{u,\eta}^N &=& a_{d,\eta}^N = -\frac12 a_{s,\eta}^N = \frac16 g_A^8 \,, \nonumber
\end{eqnarray}
with
\begin{eqnarray}
g_A^3 &=& \Delta_u^p-\Delta_d^p \,, \nonumber \\
g_A^8 &=& \Delta_u^p+\Delta_d^p-2\Delta_s^p \,. \nonumber 
\end{eqnarray}
Isospin symmetry implies $a_{u,\pi}^p = a_{d,\pi}^n$ and $a_{d,\pi}^p=a_{u,\pi}^n$.
We use the values $\Delta_u^p=\Delta_d^n=0.842$, $\Delta_d^p=\Delta_u^n=-0.427$, $\Delta_s^N=-0.085$~\cite{Hoferichter:2020osn} and neglect further dependencies on momentum transfer.

In the case of nuclei $\mathcal{N}$ instead of a nucleon, we express the nuclear matrix elements in the same way as in Eqs.~\eqref{eq:formfactors} with nuclear form factors $F_i^{f,\mathcal{N}}(q^2)$, $G_i^{f,\mathcal{N}}(q^2)$, i.e.
\begin{subequations}
\label{eq:formfactors-nucleus}
\begin{alignat}{2}
\begin{split}
& \braket{\mathcal{N}(p')|\overline{f}\gamma^\mu f |\mathcal{N}(p)} \\ &= 
\overline{u}(p')\left[F_1^{f,\mathcal{N}}(q^2)\gamma^\mu +
F_2^{f,\mathcal{N}}(q^2)\frac{i\sigma^{\mu\nu}q_\nu}{2m_\mathcal{N}}\right]u(p)\,,
\end{split} \\
\begin{split}
& \braket{\mathcal{N}(p')|\overline{f}\gamma^\mu\gamma^5 f |\mathcal{N}(p)} \\ &= 
\overline{u}(p')\left[G_A^{f,\mathcal{N}}(q^2)\gamma^\mu\gamma^5 
+G_P^{f,\mathcal{N}}(q^2)\frac{\gamma^5q^\mu}{2m_\mathcal{N}}
\right]u(p)\,.
\end{split}
\end{alignat}
\end{subequations}
For simplicity, in this work we consider  the $q^2=0$ approximation of the form factors and nuclear matrix elements of nuclei, namely $F_i^{f,\mathcal{N}}$, are obtained from up- and down-quark form factors in the same way as for the nucleons in Eq.\ (\ref{eq:istdasso?}). 
As we show below, the effect of axial couplings is strongly suppressed in the parity asymmetry parameter and nuclear weak charge. Therefore, in summary, we only use the $F_1$ form factors when considering nuclei $\carb$ and $\caes$, which gives the dominating contribution.

\subsection{Asymmetry parameter}

Concentrating on the electron-nucleus cross section induced by $\lagr_{\rm PV}$ in Eq.~\eqref{eq:eff-lagrangian} together with SM physics, i.e.\ photon and $Z$ boson exchange, we note that the amplitudes for left-handed or right-handed incoming electrons can be written as
\begin{eqnarray}
i\mathcal{M}_{L,R}^{\pm s'rr'} &=& \sum_{j=1}^{8} \frac{K_j}{\Lambda^2}\left(\overline{u}_{s'}(k_e) \ope_j u_\pm(p_e)\right) \nonumber \\
&& \times \left(\overline{u}_{r'}(k_\mathcal{N}) \ope_j' u_{r}(p_\mathcal{N})\right)\,.
\end{eqnarray}
where $K_j$, $\ope_j$, and $\ope_j'$ are given in \autoref{tab:amplitude} for the general case. In this calculation we equate  helicity and chirality of the incoming electron in order to replace $u_{\pm}(p_e)$ by $P_{R/L}u_s(p_e)$ and using trace identities from summing over $s$. The correction 
to the amplitude due to this approximation should be of order $m_e/|p_e|\approx\SI{0.5}{MeV}/\SI{155}{MeV}\approx \num{3e-3}$ and is negligible for our purposes.

\begin{table}
{\centering
\caption{Coefficients appearing in the amplitude of chiral electron-proton scattering Eq.~\eqref{eq:amplitude}.}
\label{tab:amplitude}}
\begin{tabular}{LLLL}
\toprule
j &  K_j  & \mathcal{O}_j & \mathcal{O}_j'\\
\midrule
1& C_{LVu}F_1^{u,\mathcal{N}}+C_{LVd}F_1^{d,\mathcal{N}} &
\gamma_\mu P_L & \gamma^\mu
\\
2& C_{LVu}F_2^{u,\mathcal{N}}+C_{LVd}F_2^{d,\mathcal{N}} &
\gamma_\mu P_L & {i\sigma^{\mu\nu}q_\nu}/2m_\mathcal{N}
\\
3& C_{RVu}F_1^{u,\mathcal{N}}+C_{RVd}F_1^{d,\mathcal{N}} &
\gamma_\mu P_R & \gamma^\mu
\\
4& C_{RVu}F_2^{u,\mathcal{N}}+C_{RVd}F_2^{d,\mathcal{N}} &
\gamma_\mu P_R & {i\sigma^{\mu\nu}q_\nu}/2m_\mathcal{N}
\\
5& C_{LAu}G_A^{u,\mathcal{N}} + C_{LAd}G_A^{d,\mathcal{N}} & 
\gamma_\mu P_L & \gamma^\mu\gamma^5
\\
6& C_{LAu}G_P^{u,\mathcal{N}} + C_{LAd}G_P^{d,\mathcal{N}} & 
\gamma_\mu P_L & \gamma^5q^\mu/2m_\mathcal{N}
\\
7& C_{RAu}G_A^{u,\mathcal{N}} + C_{RAd}G_A^{d,\mathcal{N}} & 
\gamma_\mu P_R & \gamma^\mu\gamma^5
\\
8& C_{RAu}G_P^{u,\mathcal{N}} + C_{RAd}G_P^{d,\mathcal{N}} & 
\gamma_\mu P_R & \gamma^5q^\mu/2m_\mathcal{N}
\\
\bottomrule
\end{tabular}
\end{table}

The differential cross section for initial polarization $X$ is proportional to the squared matrix element,
\be
\frac{\der\sigma_X}{\der t} \sim |\mathcal{M}_X|^2\,,
\ee
($t$ being the Mandelstam variable) and therefore the asymmetry parameter is simply given by the squared amplitudes in the form of
\be
A_\text{PV} = \frac{\frac{\der\sigma_R}{\der t}-\frac{\der\sigma_L}{\der t}}{\frac{\der\sigma_R}{\der t}+\frac{\der\sigma_L}{\der t}} = \frac{|\mathcal{M}_R|^2-|\mathcal{M}_L|^2}{|\mathcal{M}_R|^2+|\mathcal{M}_L|^2}\,,
\ee
where
\begin{align}
\label{eq:amplitude}
& |\mathcal{M}_{L,R}|^2 = \frac12 \sum_{j=1}^{8}\sum_{k=1}^{8}\frac{1}{\Lambda^4}K_jK_k^* \nonumber \\ 
& \times \sum_{s,s',r,r'=\pm}\text{tr}\left[
(\slashed{k}_e+m_e)\mathcal{O}_jP_{L,R}(\slashed{p}_e+m_e)P_{R,L}\gamma^0\mathcal{O}_k^\dagger\gamma^0 \right] \nonumber \\
& \times \text{tr}\left[
(\slashed{k}_\mathcal{N}+m_\mathcal{N})\mathcal{O}_j'(\slashed{p}_\mathcal{N}+m_\mathcal{N})\gamma^0\mathcal{O}_k^{\prime\dagger}\gamma^0 \right].
\end{align}
To account also for the SM contributions of photon and $Z$ boson exchange in the calculation we replace in \autoref{tab:amplitude}
\begin{subequations}
\begin{alignat}{2}
  \frac{C_{XVf}}{\Lambda^2} &\rightarrow \frac{-4\pi\alpha q_f}{q^2} + \frac{g^2}{2c_W^2}\frac{g_X^e g_V^f}{q^2-m_Z^2} + \frac{C_{XVf}}{\Lambda^2}\,,  \\
  \frac{C_{XAf}}{\Lambda^2} &\rightarrow \frac{g^2}{2c_W^2}\frac{g_X^e g_A^f}{q^2-m_Z^2} + \frac{C_{XAf}}{\Lambda^2}\,,
\end{alignat}
\end{subequations}
where the SM charges and couplings are, as usual, $g^2=4\pi\alpha/s_W^2$ (with $s_W\equiv \sin\theta_W$ being the sine of the weak mixing angle, and $\alpha=e^2/4\pi$ being the fine-structure constant), and
\be\label{eq:z-charges}
\begin{split}
q_u &= \frac23\,,\\
g_V^u &= \frac12-\frac43s_W^2\,,\\
g_V^d &= -\frac12 + \frac23 s_W^2\,,\\
g_L^e &= -\frac12 + s_W^2\,,
\end{split}
\qquad
\begin{split}
q_d &= -\frac13\,, \\
g_A^u &= \frac12\,, \\
g_A^d &= -\frac12\,, \\
g_R^e &= s_W^2\,.
\end{split}
\ee
The full resulting expression is too lengthy to reproduce here. It is, however, illustrative to re-derive certain limits from it.
To recover the leading-order SM expectation for the case of a proton, we can set all $C$ coefficients to zero such that the term of leading order in $q^2/m_Z^2$ reads
\be
\label{eq:apv-general}
A_{\text{PV}}^{\text{LO}}=
\frac{g^2}{2c_W^2}\frac{( g_R^e-g_L^e) (F_1^{u,p}(q^2) g_V^u + F_1^{d,p}(q^2) g_V^d) }{ 4\pi \alpha (F_1^{u,p} q_u+ F_1^{d,p} q_d)}\frac{q^2}{m_Z^2}\,.
\ee
Taking $F_1^{u,p}(q^2)\approx F_q^{u,p}(0)=2$ and $F_1^{d,p}(q^2)\approx F_q^{d,p}(0)=1$ we obtain the standard result~\cite{Becker:2018ggl}
\be
\label{eq:apv-proton}
A_{\text{PV}}^{\text{LO}}=
\frac{g^2}{2c_W^2}\frac{( g_R^e-g_L^e) (2 g_V^u +g_V^d) }{ 4\pi \alpha (4/3 -1/3)}
=-\frac{G_F}{\sqrt{2}} \frac{Q^2}{4\pi\alpha} (1-4 s_W^2)\,,\\
\ee
where $Q^2\equiv -q^2$. 
We observe that (weak) axial charges do not contribute at leading order in the proton case. Evaluating this expression at the central value for the scattering angle in P2, $q^2=-(\SI{93}{MeV})^2$, and for the expected low-energy value of the Weinberg angle,  $s_W^2=0.23$, results in $A_{\text{PV}}^{\text{LO}}=\num{-4.815e-8}$ for the proton. This deviates from the SM expectation $A_{\text{PV}}^{\text{pred}}=\num{-3.994e-8}$ due to radiative corrections \cite{Becker:2018ggl}. Instead of accounting for all corrections, we will consider the relative strength of deviations from the expected asymmetry due to new physics contributions $A_\text{PV}^\text{NP}$, that is, we consider $\Delta A_\text{PV}^\text{NP}/A_{\text{PV}}^\text{pred}$ as detailed below.
Before doing that, we state the leading-order asymmetry in the case of $\carb$. In this case we have, approximately, $F_1^{u,\carb}\approx 6n_{u,p}+6n_{u,n}=18$ and  $F_1^{d,\carb}\approx 6n_{d,p}+6n_{d,n}=18$, where $n_{f,N}$ denotes the number of valence quarks $f$ contained in the nucleon $N$, such that
\be
\label{eq:apv-carbon}
A_{\text{PV}}^{\text{LO}}=
\frac{g^2}{2c_W^2}\frac{( g_R^e-g_L^e) (18 g_V^u +18g_V^d) }{ 4\pi \alpha 18(2/3 -1/3)}
=\frac{G_F}{\sqrt{2}} \frac{Q^2}{4\pi\alpha} 4s_W^2\,.
\ee

We have examined possible uncertainties on the theoretical predictions for $A_{{\rm PV}}$
and identified that the largest uncertainty arises from the form factors.
The various form factors 
used in the calculation are generally $q^{2}$ dependent. The weak
interaction contribution relies on the weak charge form factor 
\begin{align}
    F_{Z}(q^{2})=F_{1}^{u,p}(q^{2})g_{V}^{u}+F_{1}^{d,p}(q^{2})g_{V}^{d}
\end{align}
(for leptoquark and other new physics, one has different combinations
of $F_{1}^{u,p}$ and $F_{1}^{d,p}$) which differs from the electric
charge form factor 
\begin{align}
    F_{\gamma}(q^{2})=F_{1}^{u,p}(q^{2})q_{u}+F_{1}^{d,p}(q^{2})q_{d} \, .
\end{align}
When $q^{2}$ is not negligibly small, the $q^{2}$ dependence of
these form factors corrects  the results as 
\begin{align}
    A_{{\rm PV}}\rightarrow A_{{\rm PV}}\frac{F_{Z}(q^{2})}{F_{Z}(0)}\frac{F_{\gamma}(0)}{F_{\gamma}(q^{2})} \, .
\end{align}
A simple approach to estimate the correction is to make use of 
\[
R_{Z/\gamma}^{2}\equiv\frac{6}{F_{Z/\gamma}(0)}\left.\frac{dF_{Z/\gamma}(q^{2})}{dq^{2}}\right|_{q^{2}\rightarrow0},
\]
which is the weak/electric charge radius of proton. Using experimentally
determined values $R_{\gamma}\approx0.84$ fm and $R_{Z}\approx1.55$
fm~\cite{Horowitz:2018yxh}, we obtain $F_{\gamma}(q^{2})/F_{\gamma}(0)\approx1-0.026$
and $F_{Z}(q^{2})/F_{Z}(0)\approx1-0.089$ for $q^{2}=-(93\ {\rm MeV})^{2}$,
and hence $A_{{\rm PV}}\rightarrow(1-0.063)A_{{\rm PV}}$, which implies
that including the $q^{2}$ dependence of form factors leads to a
$6.3\%$ smaller value of $A_{{\rm PV}}$ in the SM. As for new physics
predictions, the results generally depend on different combinations
of $F_{1}^{u,p}(q^{2})$ and $F_{1}^{d,p}(q^{2})$, and hence different
form factors. The uncertainty according to the above analysis is expected
to be at the percent level as well. For $^{12}{\rm C}$, the correction
is also at the percent level according to the Helm analytic approximation
for nuclear form factors~\cite{Helm:1956zz}. 

\subsection{SMEFT operators}

Let us now consider single-operator extensions of the SM. 
We have to carefully expand the contributions in terms of small parameters in order to extract the correct leading contribution of a given operator. 
We can broadly classify the energy scales involved into small scales, $m_e$, $m_p$, $E_e$, $q^2$, and large scales $m_Z$ and $\Lambda$. Without new physics, the results can generally be expanded in inverse powers of $m_Z$. In the presence of new physics parametrized by $C/\Lambda^2$, we can make a double expansion in powers $m_Z^{-2n}$ and $\Lambda^{-2n}$. The leading new-physics contribution, if present, should be of order $\Lambda^{-2}$. We find the following leading new-physics contributions
\begin{subequations}
\begin{alignat}{2}
\label{eq:delta-apv-EFT1}
\Delta A_\text{PV}^{LVf}(\mathcal{N}) &\approx \frac{C_{LVf}}{\Lambda^2} \frac{q^2}{4\pi\alpha}
\frac{F_1^{f,\mathcal{N}}}{q_u F_1^{u,\mathcal{N}}+q_d F_1^{d,\mathcal{N}}}\,,
\\
\label{eq:delta-apv-EFT2}
\Delta A_\text{PV}^{RVf}(\mathcal{N}) &\approx -\frac{C_{RVf}}{\Lambda^2} \frac{q^2}{4\pi\alpha}
\frac{F_1^{f,\mathcal{N}}}{q_u F_1^{u,\mathcal{N}}+q_d F_1^{d,\mathcal{N}}}\,,
\\
\begin{split}
\Delta A_\text{PV}^{LAf}(\mathcal{N}) &\approx 
\frac{C_{LAf}}{\Lambda^2} G_A^{f,\mathcal{N}}
\frac{E_e q^4}{4\pi\alpha (2 E_e^2-m_e^2) m_\mathcal{N}} \\
&\times \frac{q_u (F_1^{u,\mathcal{N}}+F_2^{u,\mathcal{N}})+q_d (F_1^{d,\mathcal{N}}+F_2^{d,\mathcal{N}})}{
(q_u F_1^{u,\mathcal{N}}+q_d F_1^{d,\mathcal{N}})^2}\,,
\end{split} \\
\begin{split}
\Delta A_\text{PV}^{RAf}(\mathcal{N}) &\approx 
\frac{C_{RAf}}{\Lambda^2} G_A^{f,\mathcal{N}}
\frac{E_e q^4}{4\pi\alpha (2 E_e^2-m_e^2)m_\mathcal{N} } \\
& \times \frac{q_u (F_1^{u,\mathcal{N}}+F_2^{u,\mathcal{N}})+q_d (F_1^{d,\mathcal{N}}+F_2^{d,\mathcal{N}})}{
(q_u F_1^{u,\mathcal{N}}+q_d F_1^{d,\mathcal{N}})^2}\,,
\label{eq:delta-apv-EFT3}
\end{split}
\end{alignat}
\end{subequations}
where $f=u,d$. This also shows that the sensitivity on axial interactions is lower due to the additional suppression by $q^2/(m_\mathcal{N}E_e)$.
One can check numerically for proton and $\carb$ that these are indeed the leading contributions to the correction to the asymmetry. However, we will use the exact expressions for our numerical results. It is not surprising that no power of $m_Z^{-2}$ is required here, because these four operators are by themselves parity-violating. The (pseudo)scalar and tensor operators, being parity-conserving, would require a cross-term with the SM-intrinsic parity violation expressed through the $Z$-couplings in order to contribute to $\Delta A_\text{PV}$. Therefore the sensitivity of P2 to those interactions is very weak.  This justifies again our separation into parity-violating interactions in Eq.~\eqref{eq:eff-lagrangian} and parity-conserving interactions in Eq.~(\ref{eq:eff-lagrangian-spt}).

The sensitivity of P2 to new physics can be estimated by the same method as in Ref.~\cite{Dev:2021otb}. Namely, we require that the new physics contribution does not exceed
\be
\label{eq:sensitivities}
\frac{\Delta A_\text{PV}}{A_\text{PV}} = 
\begin{cases}
 \sqrt{3.84}\times 1.4\% = 2.74\% &\quad \mathcal{N}=p\,,\\
 \sqrt{3.84}\times 0.3\% = 0.59\% &\quad \mathcal{N}=\text{$\carb$}\,,
\end{cases}
\ee
which corresponds to 95\% CL. These numbers correspond to the expected sensitivities of the P2  experiment according to Ref.~\cite{Becker:2018ggl}.
For the normalization in the case of protons, we use the expected SM value $A_\text{PV}^\text{pred}=\num{-3.994e-8}$. 
For the normalization in the case of $\carb$, we use the leading-order SM result from our own calculation. Moreover, we approximate $F_1^{u,\carb}=F_1^{d,\carb}=18$ while neglecting the other form factors. This leads to expected minimal values for new physics scales $\Lambda/\sqrt{C}$ given in \autoref{tab:bounds}. If we  use the replacement rules Eq.~\eqref{eq:EFT-matching-nucleon}, we can alternatively project bounds on SMEFT coefficients in the same way. The results are collected in \autoref{tab:bounds-smeft} along with existing bounds from APV and ATLAS dilepton searches discussed in \autoref{sec:other}. 

\begin{table}
{\centering
\caption{Expected bounds at 95\%~confidence level (CL) on the scale $\Lambda/\sqrt{C}$ of the operators in Eq.\  \eqref{eq:EFT-matching-nucleon} from P2 for the case of proton and carbon targets, assuming a single coupling at a time.}
\label{tab:bounds}}
\begin{tabular}{LLLLL}
\toprule
\text{Target}& C & \Lambda/\sqrt{C} \; \text{[TeV]} & C & \Lambda/\sqrt{C}\; \text{[TeV]}\\
\midrule
\multirow{4}{*}{$p$} &C_{LVu} & 13.1 & C_{LVd} & 9.3\\
&C_{RVu} & 13.1 & C_{LVd} & 9.3 \\
&C_{LAu} & 2.6 & C_{LAd}  & 1.8\\
&C_{RAu} & 2.6 & C_{LAd} &  1.8 \\
\midrule
\multirow{2}{*}{$\carb$} & C_{LVu} & 8.4 & C_{LVd} & 8.4\\
&C_{RVu} & 8.4 & C_{RVd} & 8.4 \\
\bottomrule
\end{tabular}
\end{table}

\begin{table}
{\centering
\caption{Expected bounds at 95\%~CL on the scale $\Lambda/\sqrt{C}$ of SMEFT operators  from P2 for the case of proton and carbon targets, assuming a single coupling at a time. These are compared to bounds from APV~\cite{Sahoo:2021thl} in \autoref{sec:APV} and  ATLAS dilepton limits~\cite{ATLAS:2020yat} in \autoref{sec:collider}.}
\label{tab:bounds-smeft}}
\begin{tabular}{LLLLL}
\toprule
\multirow{2}{*}{\text{Coupling}}   
&\multicolumn{4}{L}{\Lambda/\sqrt{C}\; \text{[TeV]}}\\
\addlinespace[4pt]
& \text{ P2 ($p$)}
&  \text{ P2 ($\carb$)}
& \text{ APV ($\caes$)}
& \text{ ATLAS dilepton}\\
\midrule
C_{lq(1)}  
&11.4 & 8.4 & 11.1 & 7.7\\
C_{lq(3)}  
& 6.9 & \text{---} & 2.6 & 9.2\\
C_{eu}
& 9.4 & 5.9 & 3.2 & 7.5\\
C_{ed}
& 6.4 & 5.9 & 3.4 & 4.8\\
C_{qe}
& 11.3 & 8.4 & 4.6 & 7.2\\
C_{lu}
&9.0 & 5.9 & 7.6 & 6.2\\
C_{ld}
& 6.7 & 5.9 & 8.1 & 5.0\\
\bottomrule
\end{tabular}
\end{table}

Here we would like to compare our results with a recent analysis 
in Ref.~\cite{Crivellin:2021bkd} which gives P2 bounds in terms of the operators
$\frac{G_{F}}{\sqrt{2}}C_{1q}^{e}\overline{q}\gamma^{\mu}q\overline{e}\gamma_{\mu}\gamma_{5}e$.
The 2$\sigma$ widths of $C_{1u}^{e}$ and $C_{1d}^{e}$ from Figs.~1-3
in Ref.~\cite{Crivellin:2021bkd} are $6.46\times10^{-4}$ and $1.33\times10^{-3}$
respectively. Recasting to our $C_{XVu}$ and $C_{XVd}$ ($X=L$ or $R$) 
Wilson coefficients, they correspond to $\Lambda/\sqrt{C}=13.7$
TeV and $9.6$ TeV, which are consistent with the first two rows in
Tab.~III. Note that Ref.~\cite{Crivellin:2021bkd} did not consider axial-vector operators nor the $^{12}$C case.

\subsection{Leptoquarks}
 
\begin{figure*}
\centering
\includegraphics[scale=0.6]{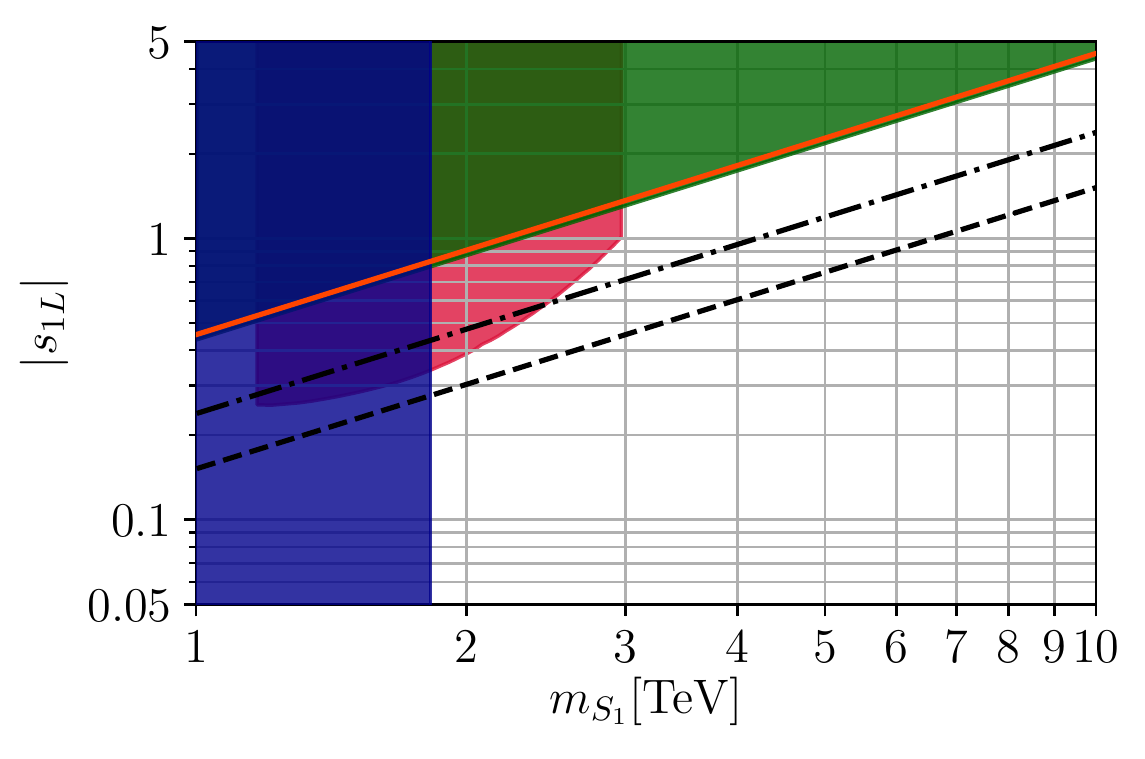}
\includegraphics[scale=0.6]{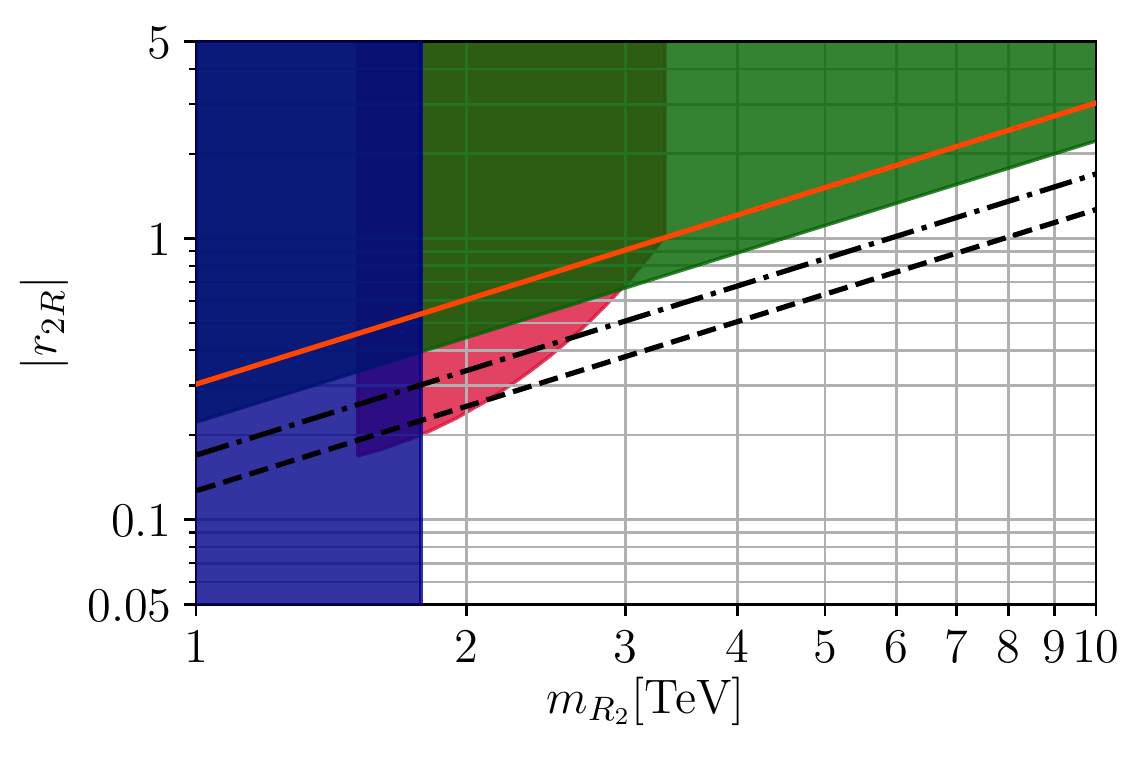}\\
\includegraphics[scale=0.6]{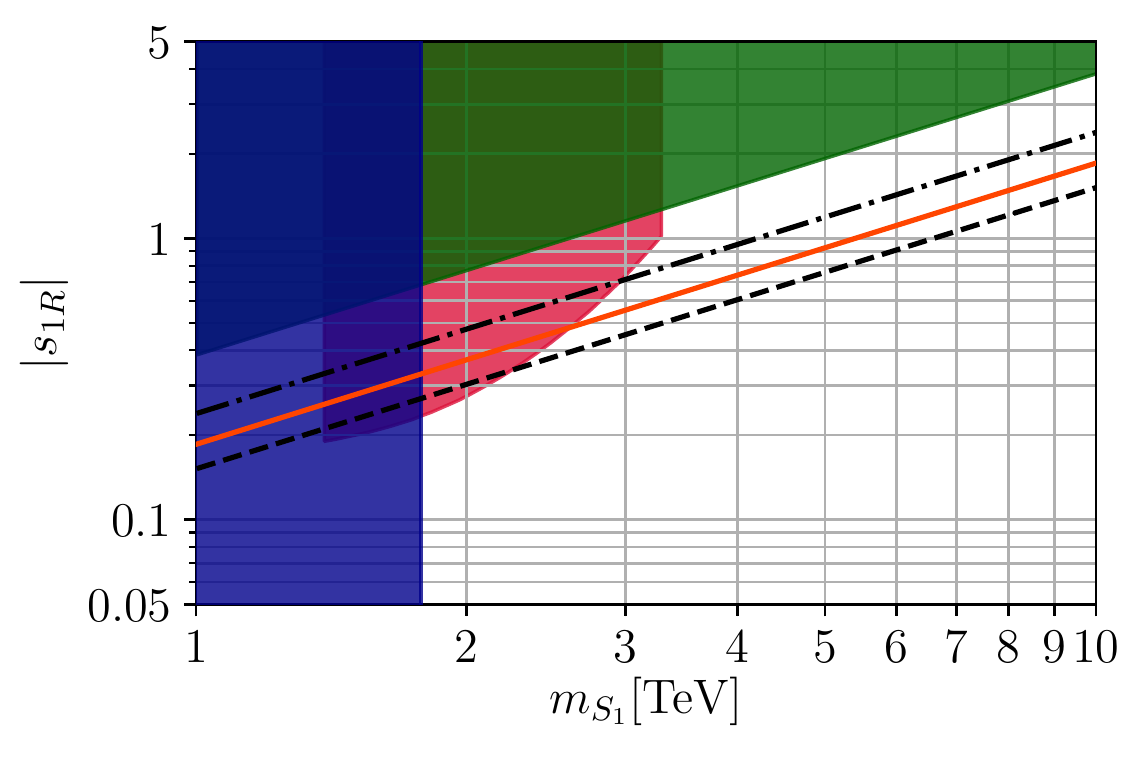}
\includegraphics[scale=0.6]{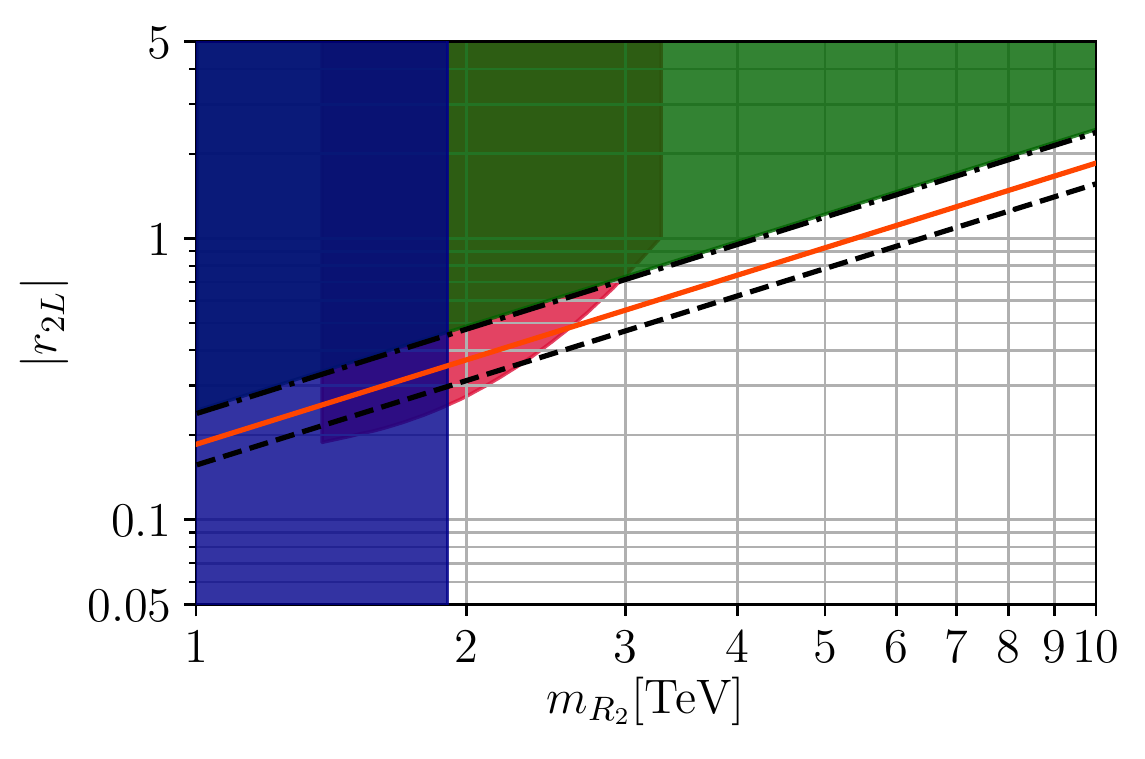}\\
\includegraphics[scale=0.6]{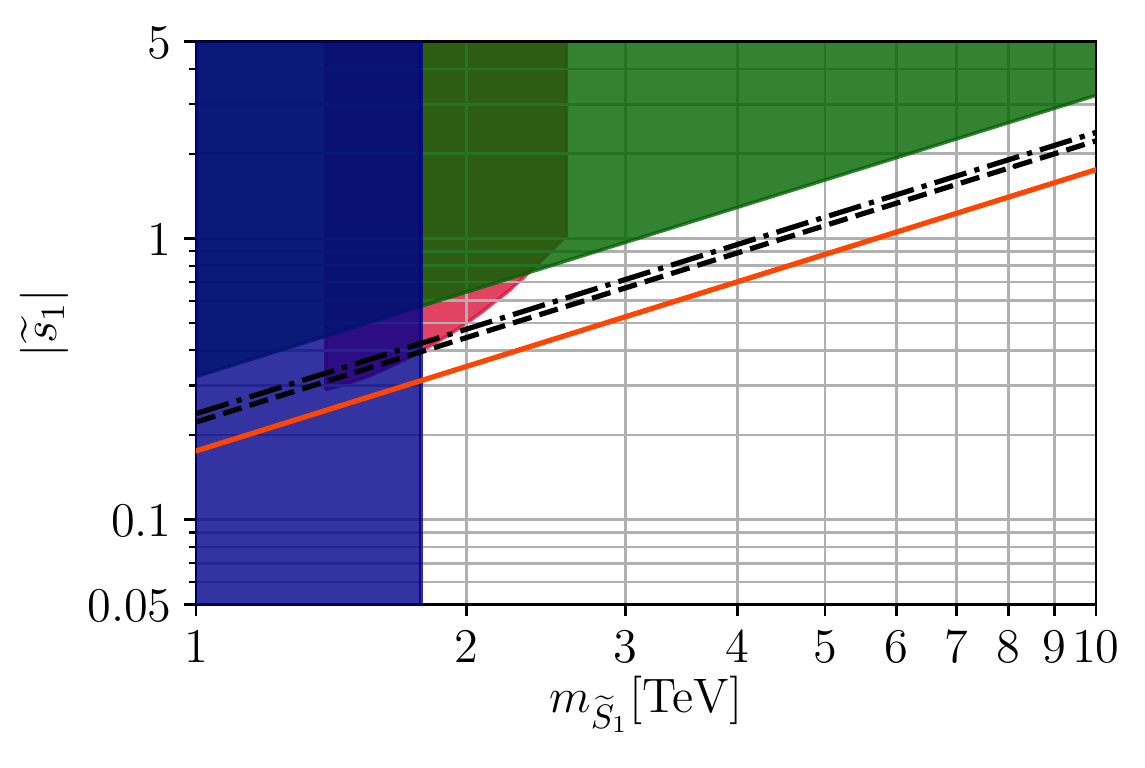}
\includegraphics[scale=0.6]{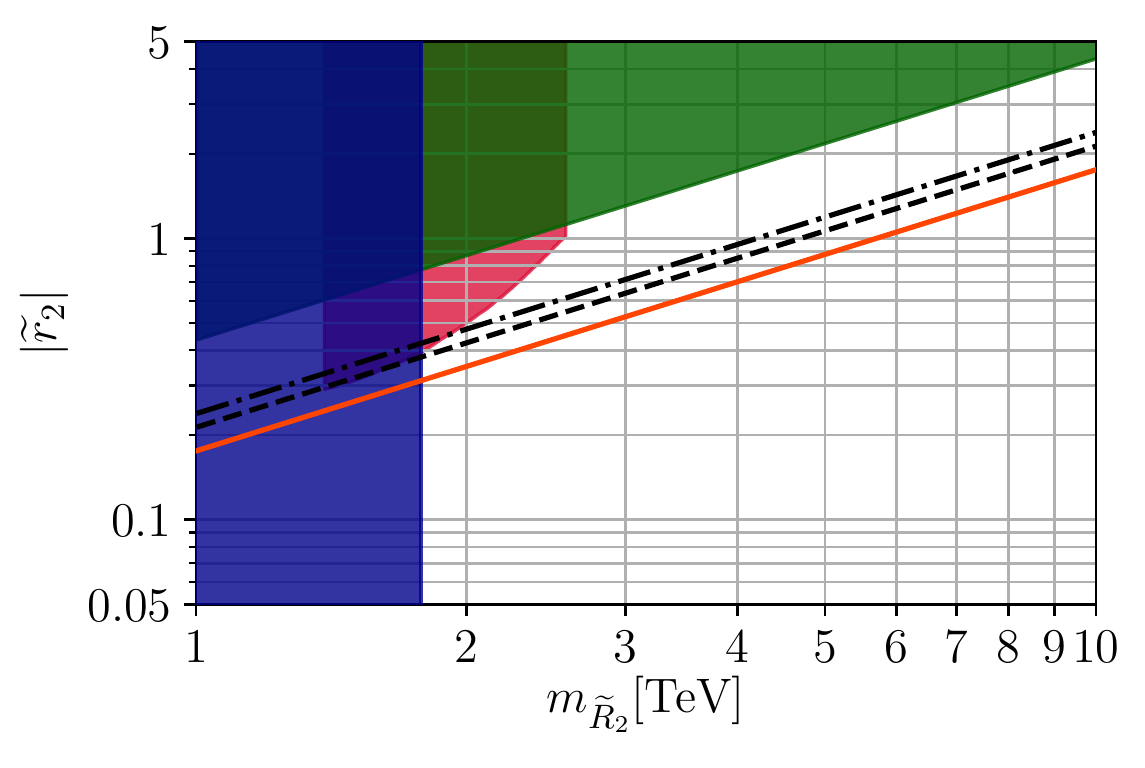}\\
\includegraphics[scale=0.6]{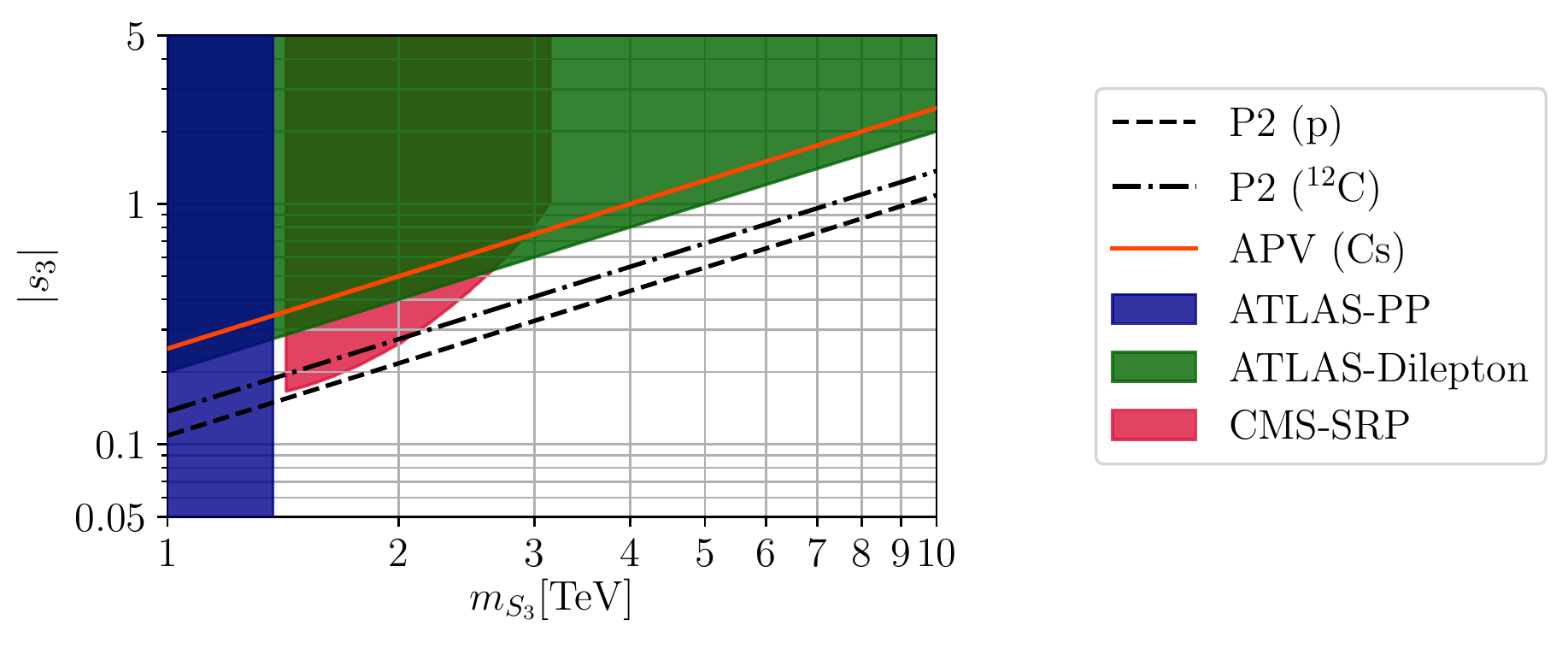}
\caption{Expected 95\% CL exclusion limits from P2 for scalar leptoquarks assuming one coupling to dominate, compared to existing bounds from APV~\cite{Sahoo:2021thl} and the LHC limits from  pair production (PP)~\cite{ATLAS:2020dsk, Diaz:2017lit}, dilepton~\cite{ATLAS:2020yat} and single resonant production (SRP)~\cite{Crivellin:2021egp} channels. The constant ratios $m_\text{LQ}/g_\text{LQ}$ from P2, APV and ATLAS-Dilepton correspond to the entries of \autoref{tab:lq-bounds}. }
\label{fig:lqbounds-scalar}
\end{figure*}

\begin{figure*}
\includegraphics[scale=0.6]{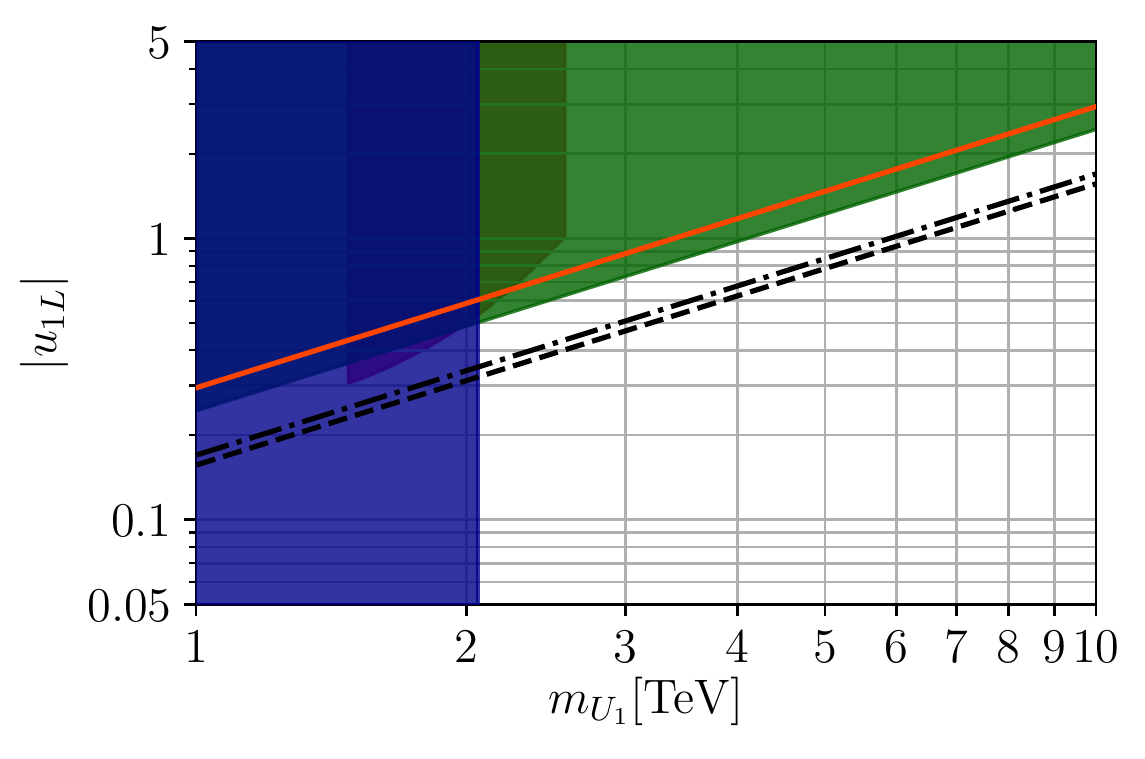}
\includegraphics[scale=0.6]{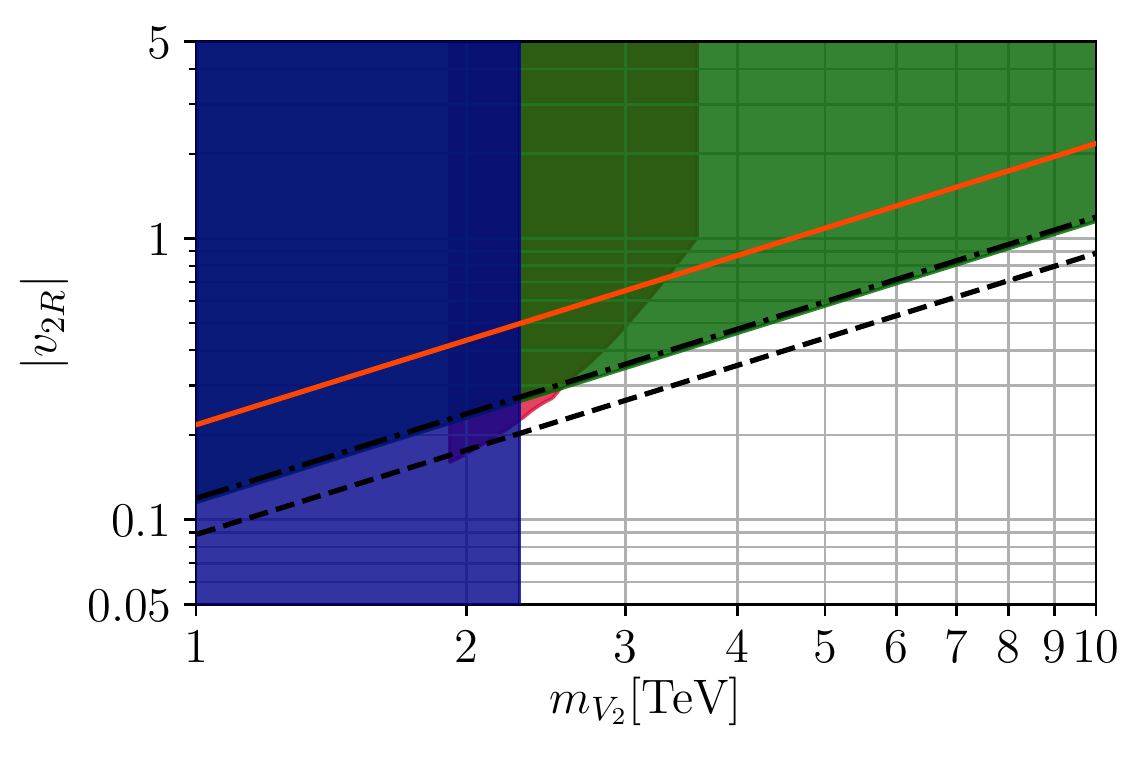}\\
\includegraphics[scale=0.6]{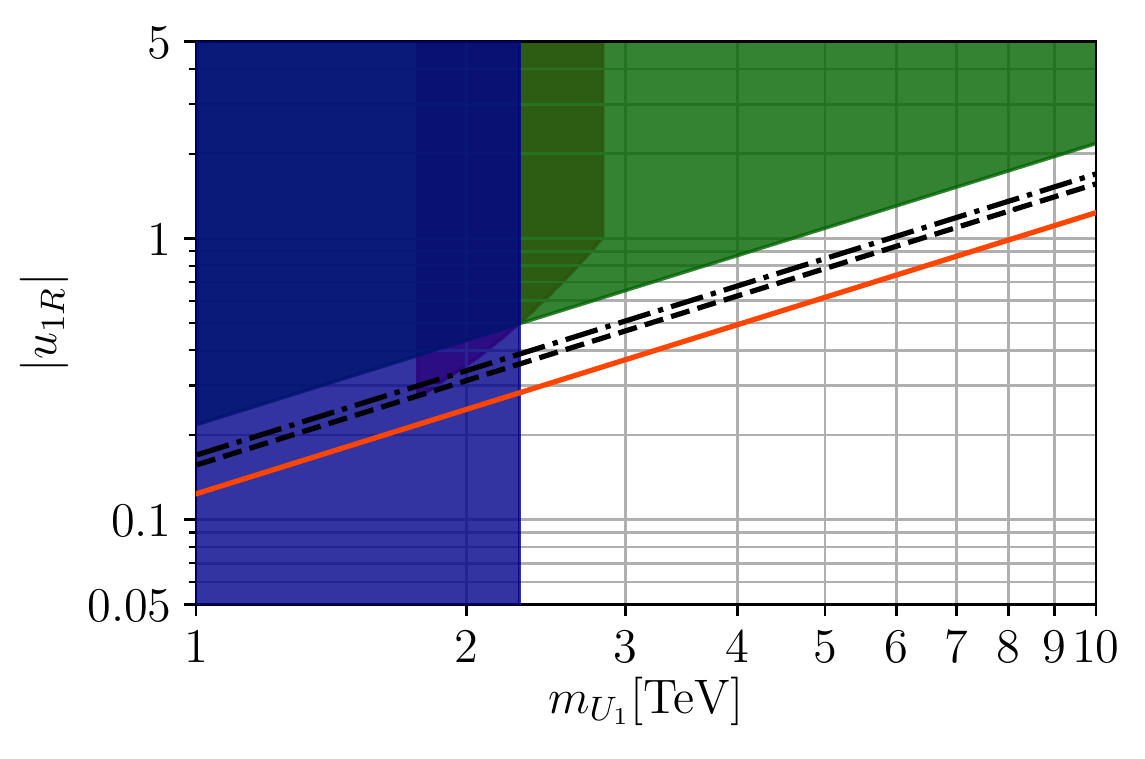}
\includegraphics[scale=0.6]{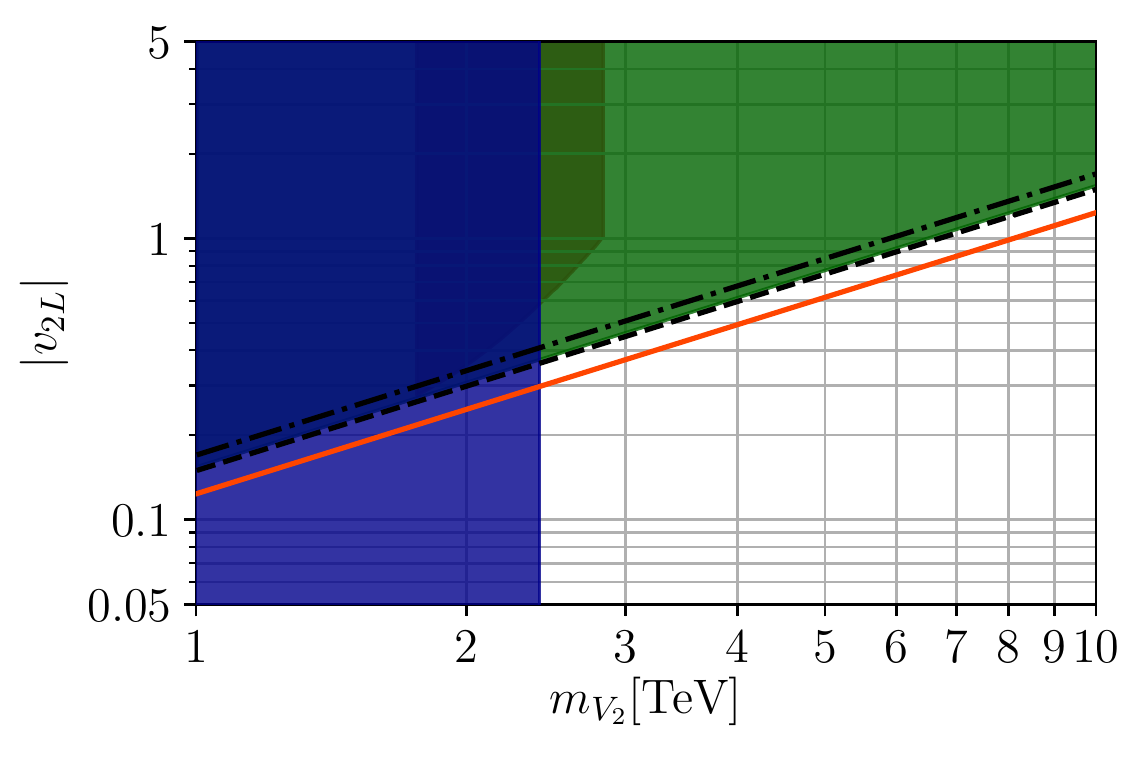}\\
\includegraphics[scale=0.6]{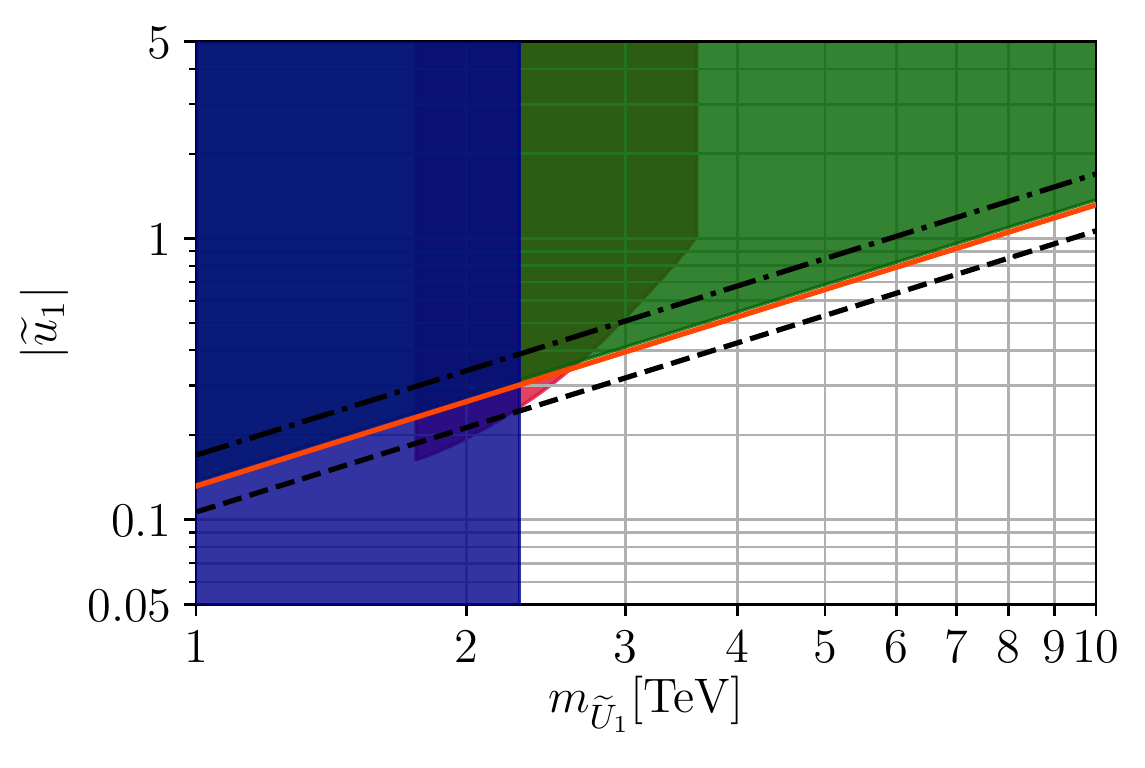}
\includegraphics[scale=0.6]{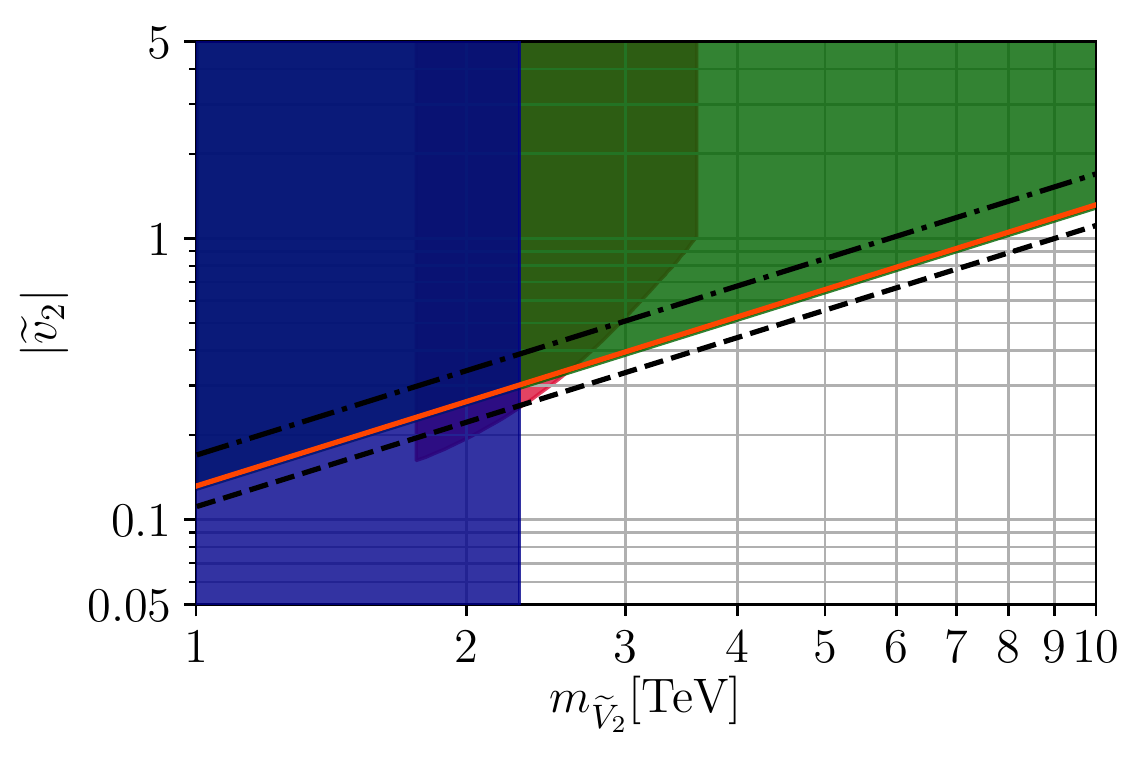}\\
\includegraphics[scale=0.6]{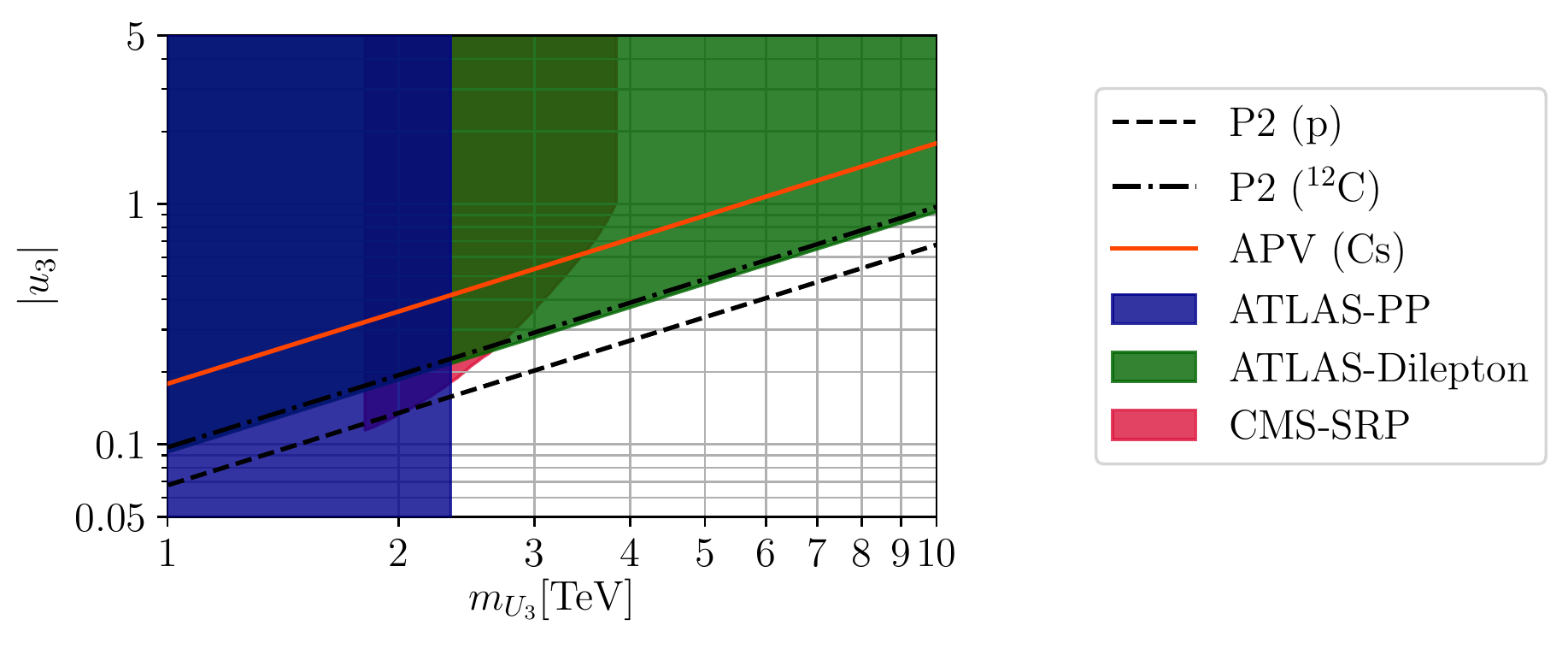}
\caption{Expected 95\% CL exclusion limits from P2 for vector leptoquarks assuming one coupling to dominate, compared to existing bounds from APV~\cite{Sahoo:2021thl} and the LHC limits from pair production (PP)~\cite{ATLAS:2020dsk, Diaz:2017lit}, dilepton~\cite{ATLAS:2020yat} and single resonant production (SRP)~\cite{Crivellin:2021egp} channels. The constant ratios $m_\text{LQ}/g_\text{LQ}$ from P2, APV and ATLAS-Dilepton correspond to the entries of \autoref{tab:lq-bounds}. }
\label{fig:lqbounds-vector}
\end{figure*}

As obvious from Eqs.\ (\ref{eq:lq-couplings-first})-(\ref{eq:lq-couplings-last}), the same SMEFT operator is generated by different leptoquarks.  
 If we assume one coupling to dominate, we can constrain $m_\text{LQ}/g_\text{LQ}$ in the EFT-like limit of $m_\text{LQ} \gg q^2$. To achieve this, one can take the expression for the asymmetry in terms of the coefficients in Eq.~\eqref{eq:EFT-matching-nucleon} and then use the matching of leptoquark couplings to SMEFT operators in Eqs.~\eqref{eq:lq-couplings-first}-\eqref{eq:lq-couplings-last}. The results are 
 shown in \autoref{tab:lq-bounds} as well as Figures~\ref{fig:lqbounds-scalar}-\ref{fig:lqbounds-vector}.
 Since for $\carb$ we are neglecting momentum dependence of the form factors as well as the contributions of leptoquarks to axial currents (which are suppressed, as discussed below Eq.~\eqref{eq:delta-apv-EFT3}), those numbers represent the leading order with respect to form factor and momentum-dependent corrections. They are, however, sufficient to compare sensitivities of different targets to different interactions. 
 
 We observe that generically the bounds using the proton target are stronger than the ones using $^{12}$C. 
 This can be explained by the smaller $A_\text{PV}^\text{LO}$ of the proton, which is proportional to $(1-4s_W^2)\approx0.08$ compared to $(4s_W^2)\approx0.92$ for $\carb$. Considering that $\Delta A_\text{PV}$ in Eqs.~\eqref{eq:delta-apv-EFT1}~and~\eqref{eq:delta-apv-EFT2} for up quarks and down quarks differs only by the factors
\begin{subequations}
\begin{alignat}{2}
\frac{F_1^{f,p}}{q_u F_1^{u,p}+q_d F_1^{d,p}} &\approx \begin{cases} 2 & f=u\\ 1 & f=d\end{cases}\,, \\
\frac{F_1^{f,\carb}}{q_u F_1^{u,\carb}+q_d F_1^{d,\carb}} &\approx \begin{cases} 3 & f=u\\ 3 & f=d\end{cases}\,,
\end{alignat}
\end{subequations}
and using the expected sensitivities in Eq.~\eqref{eq:sensitivities}, we can estimate that the constraints on $C/\Lambda^2$ from $p$ should be stronger by a factor of approximately
 \be
 \frac{0.92\cdot 0.3\%}{0.08\cdot 1.4\%}\cdot\begin{cases} \frac{2}{3} & f=u \\ \frac13 & f=d \end{cases} \ 
 \approx \ \begin{cases} 1.7 & f=u \\ 0.8 & f=d \end{cases}\, 
 \ee
 compared to $\carb$. This explains why couplings to down quarks are constrained to similar magnitude, while couplings to up quarks are better constrained by proton targets. These factors, however, are not exact, since we did not take radiative and form factor corrections into account.

\begin{table}
{\centering
\caption{Expected bounds on leptoquarks masses at 95\%~CL for single couplings $g_\text{LQ}=1$ from P2 compared with existing bounds from APV~\cite{Sahoo:2021thl} and ATLAS dilepton data~\cite{ATLAS:2020yat}. Additional bounds are shown in Figures~\ref{fig:lqbounds-scalar}~and~\ref{fig:lqbounds-vector}.}
\label{tab:lq-bounds}}
\begin{tabular}{LLLLLL}
\toprule
\multirow{2}{*}{\text{Leptoquark}}
 & \multirow{2}{*}{\text{Coupling}}   
&\multicolumn{4}{c}{$m_\text{LQ}\text{[TeV]}$}\\
\addlinespace[4pt]
&& \text{ P2 ($p$)}
&  \text{ P2 ($\carb$)}
& \makecell{ \text{APV} \\ \text{($\caes$)} }
& \makecell{ \text{ATLAS} \\ \text{dilepton} }\\
\midrule
S_1 & s_{1L}  
&6.6 & 4.2 & 2.2 & 2.3\\
S_1 & s_{1R}   
&6.6 & 4.2 & 5.4 & 2.6\\
\widetilde{S}_1 & \widetilde{s}_1    
&4.5 & 4.2 & 5.7 & 3.1\\
S_3 & s_3   
& 9.2 & 7.3 & 4.0 & 5.0 \\
V_2 & v_{2R}
& 11.3 & 8.4 & 4.6 & 8.7\\
V_2 & v_{2L}
& 6.7 & 5.9 & 8.1 & 6.5\\
\widetilde{V}_2 & \widetilde{v}_{2}
& 9.0 & 5.9 & 7.6 & 7.8\\
R_2 & r_{2R} 
& 7.9 & 5.9 & 3.3 &4.5\\
R_2 & r_{2L} 
& 6.4 & 4.2 & 5.4 & 4.1\\
\widetilde{R}_2 & \widetilde{r}_{2}
& 4.7 & 4.2 & 5.7 & 2.3\\
U_1 & u_{1L} 
& 6.4 & 5.9 & 3.4 & 4.1\\
U_1 & u_{1R}
& 6.4 & 5.9 & 8.1 & 4.6\\
\widetilde{U}_1 & \widetilde{u}_{1}
& 9.4 & 5.9 & 7.6 & 7.3\\
U_3 & u_{3} 
& 14.8 & 10.3 & 5.6 & 10.8
\\
\bottomrule
\end{tabular}
\end{table}

 
\section{Other observables}
\label{sec:other}

A number of different observables are suitable to constrain the relevant SMEFT operators or leptoquarks. Besides electron scattering, APV probes the same couplings. Additionally, we include bounds from the LHC on leptoquark production, as well as their contribution to Drell-Yan production of electron pairs.
The operators involving lepton doublets can further be probed by coherent elastic neutrino-nucleus scattering (CE$\nu$NS), since they lead to analogous interactions of neutrinos with nuclei. However, currently CE$\nu$NS is not competitive with other constraints in this respect~\cite{Crivellin:2021bkd}. 
The current order of magnitude can be estimated by noting that the best constraints on neutrino Non-Standard Interactions (NSI) from CE$\nu$NS are at the order of $\epsilon_{V}\lsim 0.5$ for vector interactions, while bounds on axial interactions are weaker~\cite{Papoulias:2019txv,CONUS:2021dwh}. Using the mapping between NSI and SMEFT operators e.g.\ in Ref.~\cite{Bischer:2019ttk}, this can be translated to EFT scales of 
\be 
\frac{\Lambda}{\sqrt{C}} \lsim \left(\frac{G_F\epsilon_V}{\sqrt{2}}\right)^{-1/2} \approx \SI{0.5}{TeV}\,,
\ee
which are well below the values in \autoref{tab:bounds-smeft}. 
Therefore, we focus on the constraints which are most competitive with P2, namely those from APV and the LHC.

\subsection{Atomic parity violation}
\label{sec:APV}

As in Ref.\ \cite{Dev:2021otb}, we use the most precise measurement of APV which concerns the $6S_{1/2}$-$7S_{1/2}$ nuclear transition in $^{133}$Cs.
Following Ref.~\cite{Becker:2018ggl}, we define the proton weak charge $Q_W(p)$ as the limit of the asymmetry at zero-momentum transfer, normalized such that the asymmetry formula (recall that $Q^2=-q^2$)
\be
A_{\text{PV}}^{\text{LO}}=
-\frac{G_F}{\sqrt{2}}\frac{Q^2}{4\pi\alpha} Q_W(p)=\frac{G_F}{\sqrt{2}}\frac{q^2}{4\pi\alpha} Q_W(p)\,
\ee
holds, for which one can find from Eq.~\eqref{eq:apv-proton} that at leading-order the weak charge reads
\be
Q_W(p)=1-4s_W^2\,.
\ee
Generalizing to nuclei, we simplify Eq.~\eqref{eq:apv-general} to the form
\be
A_\text{PV}^{\text{LO}} = \frac{G_F}{\sqrt{2}}\frac{q^2}{4\pi\alpha} \frac{Q_W(\mathcal{N})}{F_1^{u,\mathcal{N}}q_u + F_1^{d,\mathcal{N}}q_d}
\ee
with
\begin{eqnarray}
Q_W(\mathcal{N}) &=& 2\left(F_1^{u,\mathcal{N}}g_V^u + F_1^{d,\mathcal{N}}g_V^d \right) \nonumber \\
&\approx& Z(\mathcal{N}) (1 - 4s_W^2) - N(\mathcal{N})\,,
\end{eqnarray}
where $Z$ and $N$ denote the nuclear charge and number of neutrons, see also  Ref.~\cite{Kumar:2013yoa},
such that
\begin{subequations}
\begin{alignat}{2}
Q_W(\carb) &\approx -24s_W^2 \,,  \\
Q_W(\caes) &\approx -23-220s_W^2 \approx -73.6 \,
\end{alignat}
\end{subequations}
consistent with Eq.~\eqref{eq:apv-carbon}. 
For $\caes$, which has the currently best-measured asymmetry, we have used $Z_\text{Cs} = 55$ and $N_\text{Cs} = 78$. 
The SM prediction including radiative corrections and the measured values are  
given by~\cite{Sahoo:2021thl} 
\begin{subequations}
\begin{alignat}{2}
\label{eq:disc}
Q_W^\text{SM}(^{133}\text{Cs}) &= -\SI{73.23(1)}{} \,, \\
Q_W^\text{exp}(^{133}\text{Cs}) &= -\SI{73.71(35)}{}\,.
\end{alignat}
\end{subequations}
These two numbers are consistent within $2\sigma$. We will use them nevertheless to illustrate how P2 could confirm or resolve the mild tension \cite{Sahoo:2021thl} between theory and experiment. 
From Eqs.~\eqref{eq:delta-apv-EFT1}-\eqref{eq:delta-apv-EFT3} we can see how the measured value would be changed from the SM prediction through additional contributions to interactions at quark level at leading order.
Note that therefore P2 data can be used to further investigate this deviation.  
We do so in \autoref{sec:deg}, finding that P2 should be able to rule out the necessary quark-level couplings at 95\% CL. 
We can write, for generic modifications $\Delta Q_W^\text{NP}$ of the weak charge,
\begin{align}
A_\text{PV} &= A_\text{PV}^\text{SM}+\Delta A_\text{PV}^\text{NP} \nonumber \\ 
&= \frac{G_F}{\sqrt{2}}\frac{q^2}{4\pi\alpha} \frac{Q_W^\text{SM}(\mathcal{N})+\Delta Q_W^\text{NP}(\mathcal{N})}{F_1^{u,\mathcal{N}}q_u + F_1^{d,\mathcal{N}}q_d} \,, 
\end{align}
with
\begin{align}
\Delta Q_W^\text{NP}(\mathcal{N}) = \Delta A_\text{PV}^\text{NP} \left(\frac{\sqrt{2}}{G_F}\frac{4\pi\alpha}{q^2}(F_1^{u,\mathcal{N}}q_u + F_1^{d,\mathcal{N}}q_d)
\right). \nonumber 
\end{align}
In the limit $q^2\rightarrow 0$ we conclude that, in the EFT picture,
\begin{eqnarray}
\Delta Q_W^\text{NP}(\caes) & = & \frac{\sqrt{2}}{G_F}\frac{1}{\Lambda^2}\left(F_1^{u,\caes}(C_{LVu}-C_{RVu}) \right. \nonumber  \\ 
&& + \left. F_1^{d,\caes}(C_{LVd}-C_{RVd})\right).
\end{eqnarray}
The momentum transfer in APV experiments is not exactly vanishing, but instead of the order of the inverse nuclear radius $Q\sim 1/r_0\sim \SI{30}{MeV}$ according to Refs.~\cite{Milstein:2001hi,Milstein:2002ai}. Other references set the momentum transfer at $Q\sim \SI{2.4}{MeV}$ which we will adapt here when considering leptoquark propagators~\cite{Safronova:2017xyt}. However, even with such non-vanishing $Q$ the axial couplings will be poorly probed, since they contribute to $\Delta Q_W^\text{NP}$ only with a suppression factor of $Q/m_\mathcal{N}\sim 10^{-5}$.
To recast the bounds on $Q_W$ into 95\% CL bounds on new physics, we will hence require that $Q_W^\text{SM}(\caes)+\Delta Q_W^\text{NP}(\caes)$ does not deviate from $Q_W^\text{exp}$ by more than $\sqrt{3.84}$ standard deviations, neglecting the error of the SM prediction.

The results are included in Figures~\ref{fig:lqbounds-scalar}-\ref{fig:lqbounds-vector}. It is interesting to note that we can identify two cases: Some leptoquarks increase $Q_W$ while others decrease it. Now since the SM expectation is already {\it above} the measured value, there is less room for new physics increasing $Q_W$ and more room for new physics decreasing $Q_W$. Therefore, the general pattern is that P2 is expected to improve bounds on couplings decreasing $Q_W$ while bounds on couplings increasing $Q_W$ will likely remain better tested by APV.

\subsection{Collider searches}
\label{sec:collider}

Effective interactions between two quarks and two electrons as parametrized by the SMEFT operators of Eq.~\eqref{eq:formalism-charged-fermion} and as induced, for instance, by heavy leptoquarks according to the matching in Eqs.~\eqref{eq:lq-couplings-first}-\eqref{eq:lq-couplings-last} can be tested at the LHC by searching for deviations from the SM in the Drell-Yan (DY) process $pp\rightarrow e^+e^-$. In the case of leptoquarks, this is mediated by a $t$-channel exchange of a leptoquark annihilating a quark-antiquark pair and producing an electron-positron pair.
The latest results from the ATLAS experiment~\cite{ATLAS:2020yat}\footnote{The corresponding CMS limits~\cite{CMS:2018nlk} are weaker, so we only consider the ATLAS results.} are given in terms of limits on the contribution of new physics to the cross section in the signal region for the cases of constructive and destructive interference. 
To quantify constraints implied by these limits, we simulate the expected new-physics contributions to the cross section in the signal regions of invariant masses of the dilepton system [2200,\,6000]\,\si{GeV} (constructive interference) and [2770,\,6000]\,\si{GeV} (destructive interference) using \textsc{MadGraph5}~\cite{Alwall:2014hca}. As further cuts we apply $p_T>\SI{30}{GeV}$ and $|\eta|<2.47$ on the final state leptons, to approximate the ATLAS specifications. We use the same parton distribution function as specified in the ATLAS analysis, namely \textsc{NNPDF23LO} with the \textsc{lhapdf} identifier 247000. To create UFO model files, we use \textsc{FeynRules}~\cite{FeynRules1,FeynRules2} to extend the included SM model file by leptoquarks, or, to simulate SMEFT operators, by a heavy neutral vector boson with the appropriate couplings to match the considered Wilson coefficients. 

The resulting lower bounds on the new-physics scale $\Lambda$ from SMEFT operators with unit Wilson coefficients consistent with the ATLAS search are given in \autoref{tab:bounds-smeft}.
Similarly, the limits on leptoquark masses calculated for single unit Yukawa couplings are given in \autoref{tab:lq-bounds} and shown in Figures~\ref{fig:lqbounds-scalar}~and~\ref{fig:lqbounds-vector}. We find that all couplings are currently constrained by the constructive interference bounds. 
As a consistency check of our results, we can compare the limit on the NP scale $\Lambda$ with ATLAS results. Since they assume equal couplings to up and down quarks, we can directly compare the operator $\mathcal{O}_{qe}$ which corresponds to the scenario $\eta_{LR}=1$ referring to the effective Lagrangian in Equation 1 of Ref.~\cite{ATLAS:2020yat}. This operator is also induced by the $r_{2R}$ leptoquark coupling. Therefore we can map both our EFT limit and our leptoquark limit to constraints on the scale $\Lambda_\text{ATLAS}$ used in Ref.~\cite{ATLAS:2020yat} in the following way:
\begin{subequations}
\begin{alignat}{2}
\begin{split}
& \frac{4\pi}{\Lambda_\text{ATLAS}^2} = \frac{|r_{2R}|^2}{2m_{R_2}^2} \\
& \Rightarrow  \Lambda_\text{ATLAS}\geq\sqrt{8\pi}m_{R_2}=\SI{22.6}{TeV}\,,    
\end{split} \\
\begin{split}
& \frac{4\pi}{\Lambda_\text{ATLAS}^2} = \frac{C_{qe}}{\Lambda^2} \\
& \Rightarrow  \Lambda_\text{ATLAS}\geq\sqrt{4\pi}\frac{\Lambda}{\sqrt{C_{qe}}}=\SI{25.5}{TeV}\,.
\end{split}
\end{alignat}
\end{subequations}
Comparing to the ATLAS result of $\Lambda_\text{ATLAS}\geq\SI{24.7}{TeV}$ we conclude that our method produces reasonable results. These results are also comparable to those obtained in Ref.~\cite{Crivellin:2021bkd}.

The presence of leptoquarks can also be tested at proton colliders through pair production or single resonant production (SRP). While pair production through gluon fusion dominates for small Yukawa couplings $g_\text{LQ}\lsim 0.1$-$1$ since its cross section is determined by the strong gauge coupling, SRP can be relevant for larger couplings~\cite{Crivellin:2021egp}. In Figures~\ref{fig:lqbounds-scalar}~and~\ref{fig:lqbounds-vector} we show bounds from SRP calculated in Ref.~\cite{Crivellin:2021egp}.
Turning to pair production, in our cases for single leptoquark couplings to electrons and first-generation quarks, the resulting signal from the subsequent decays of the leptoquark pair is given by an $e^+$-$e^-$-pair and two jets. Assuming the lifetime of leptoquarks is short enough, this search yields a lower bound on the leptoquark mass. In Ref.~\cite{ATLAS:2020dsk} the search is done for scalar leptoquarks giving a lower limit on the mass of about \SI{1.8}{TeV} at 95\% CL for leptoquarks coupling to singlets. Following the prescription of Ref.~\cite{Diaz:2017lit}, we rescale these limits by scaling the pair production cross sections of different leptoquarks depending on their coupling types and comparing them to the exclusion curve of Ref.~\cite{ATLAS:2020dsk}. The resulting bounds are included in Figures~\ref{fig:lqbounds-scalar}~and~\ref{fig:lqbounds-vector}.

\section{Potential to resolve degeneracies using multiple targets}\label{sec:deg}

\begin{figure*}
\centering
\includegraphics[scale=0.6]{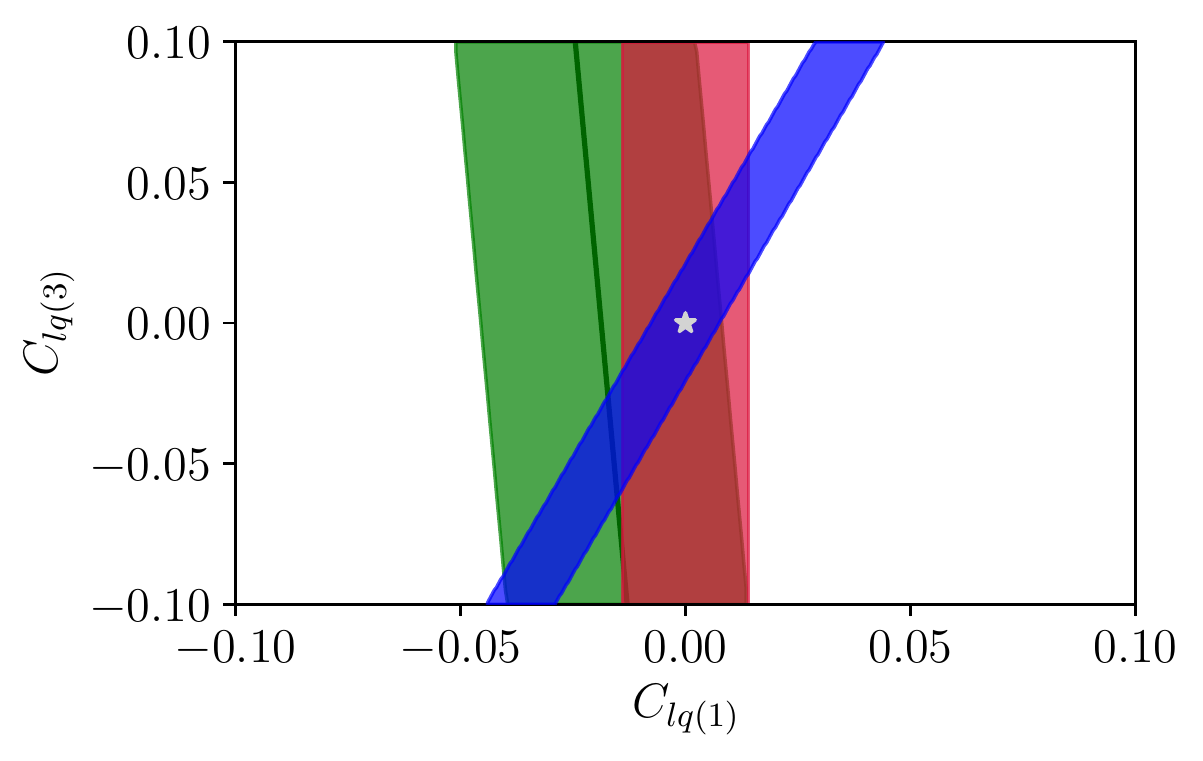}
\includegraphics[scale=0.6]{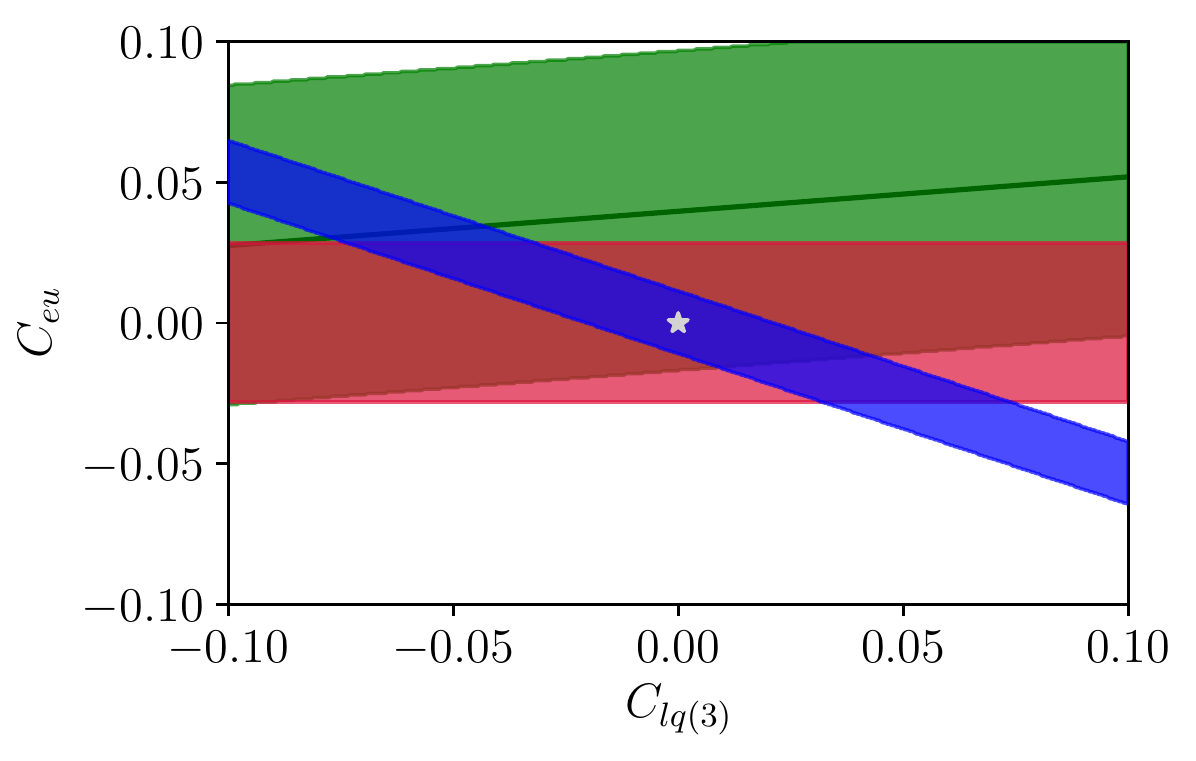}\\
\includegraphics[scale=0.6]{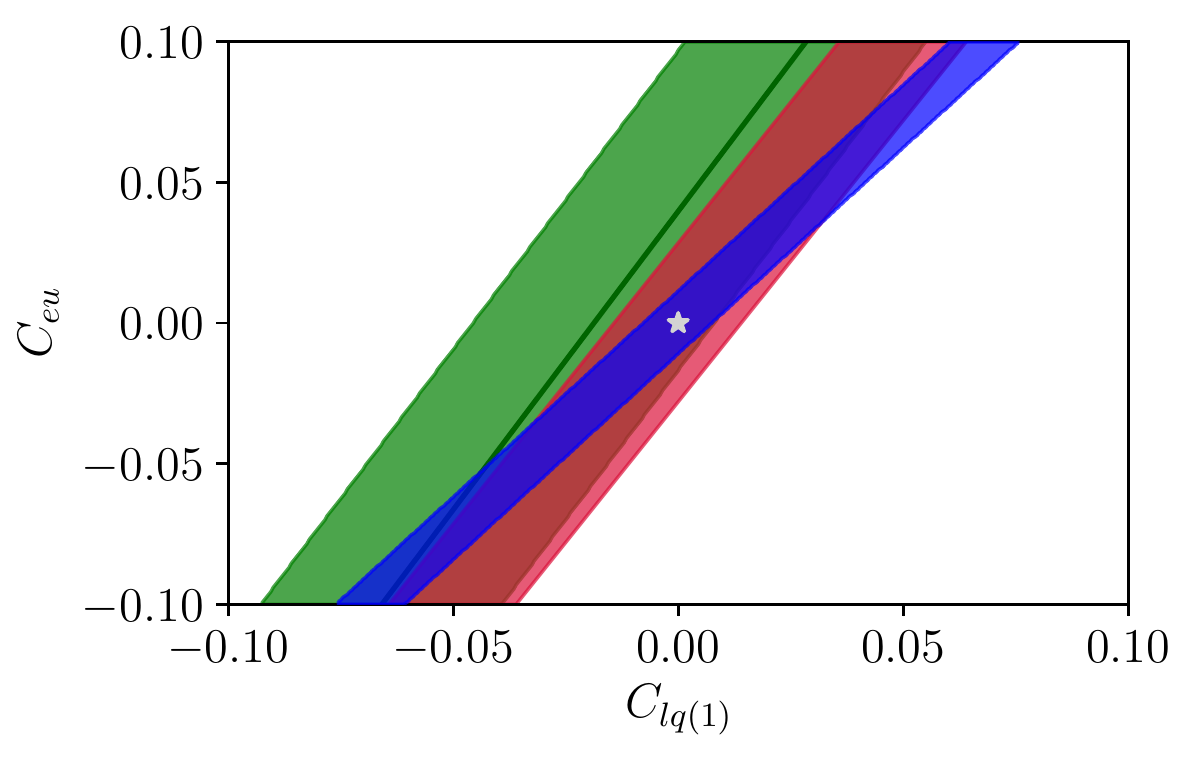}
\includegraphics[scale=0.6]{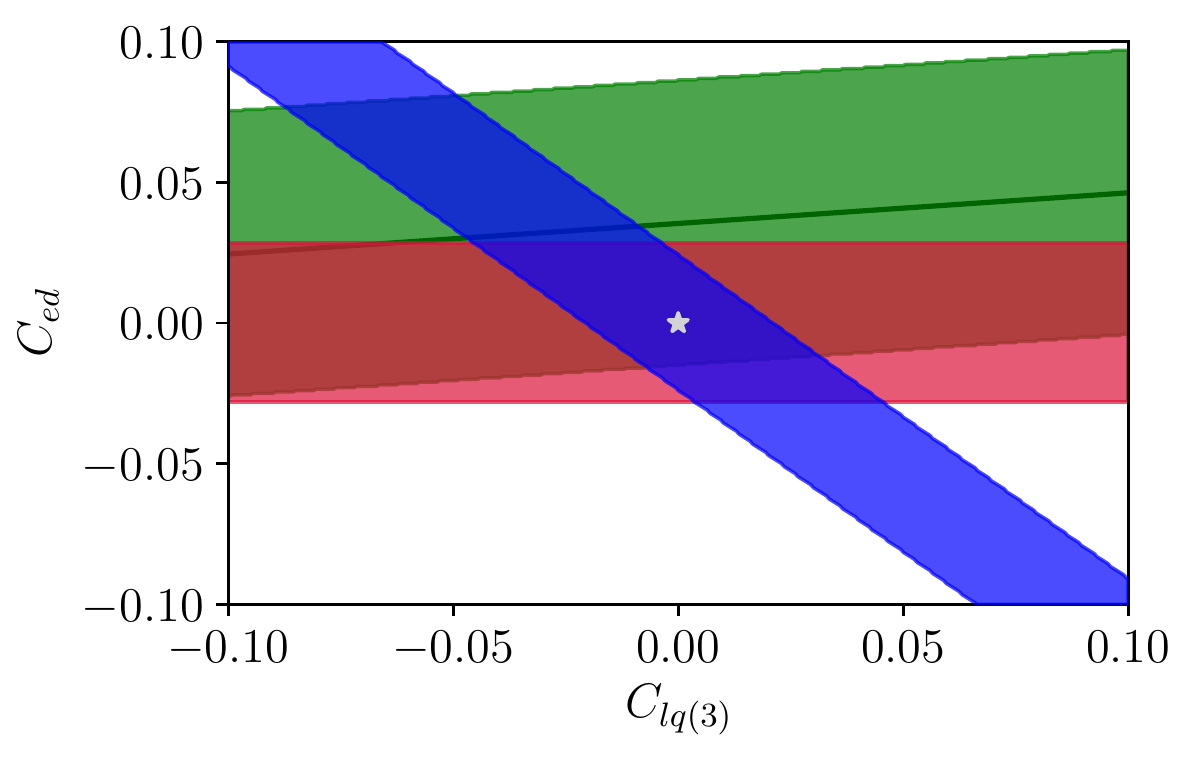}\\
\includegraphics[scale=0.6]{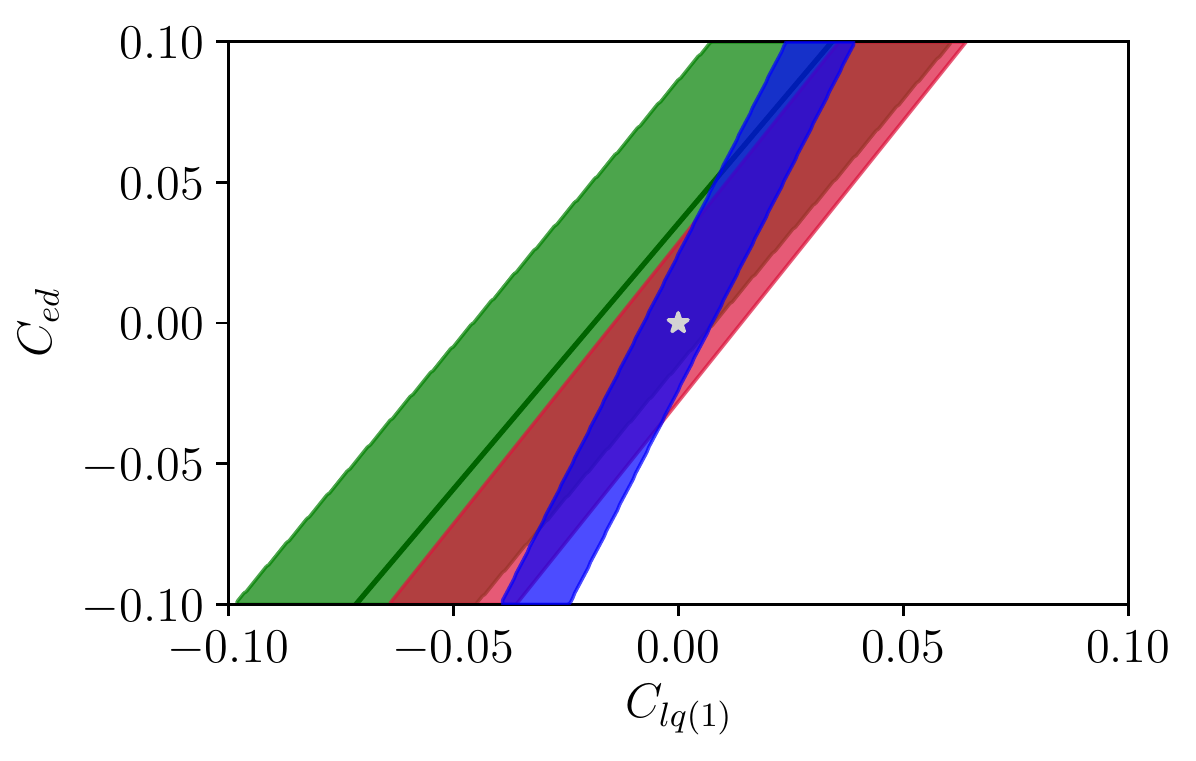}
\includegraphics[scale=0.6]{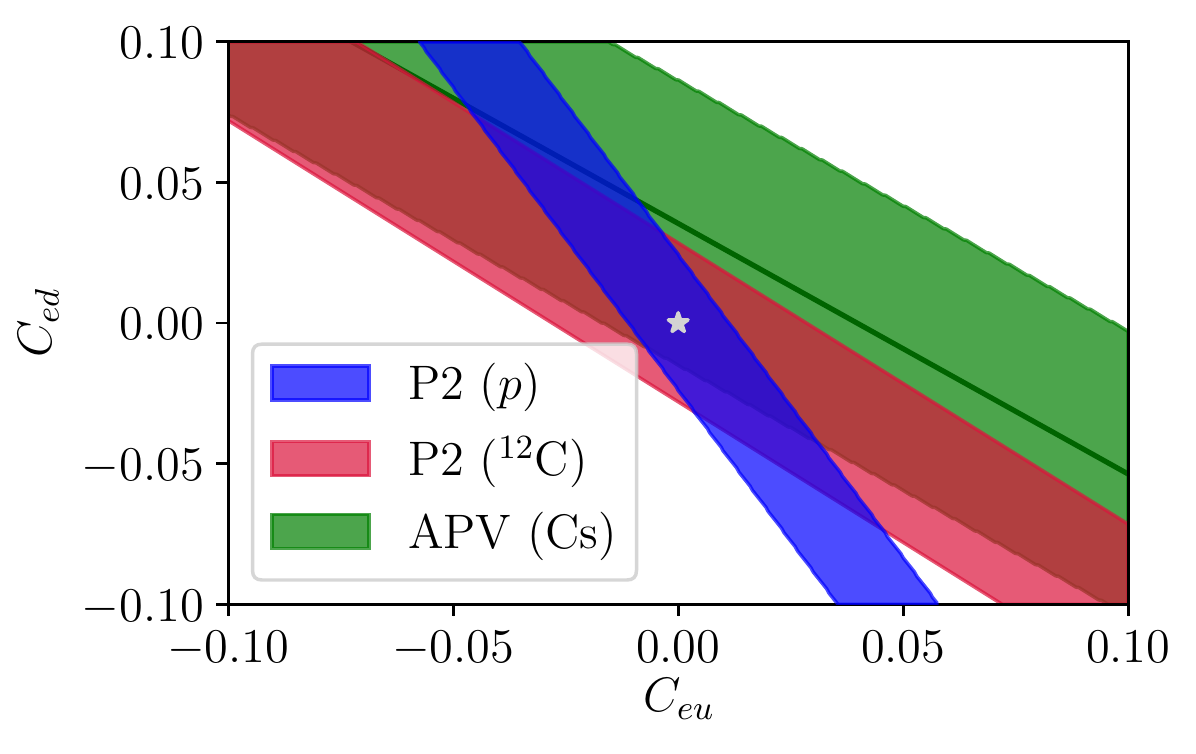}
\caption{Expected exclusion limits from P2 considering two SMEFT coefficients at a time,  compared with existing measurement from APV~\cite{Sahoo:2021thl}. The SM expectation is marked by an asterisk and the best fit to APV is shown as a dark green line. The EFT scale $\Lambda$ is set to \SI{1}{TeV}.}
\label{fig:smeftbounds}
\end{figure*}

In \autoref{fig:smeftbounds}, we show for two SMEFT coefficients at a time how the different measurements constrain the combination of parameters and break degeneracies single measurements suffer from. While the seven 
parity-violating Wilson coefficients appearing in Eq.~\eqref{eq:EFT-matching-nucleon} can be combined in 21 different ways, we only show some of the combinations, since many are basically equivalent. This is because the effective contribution to the asymmetry parameter for protons and nuclei is mainly controlled by the two effective up quark and down quark coupling coefficients
\begin{subequations}
\begin{alignat}{2}
\begin{split}
& C_{\text{PV}u} = C_{LVu}-C_{RVu} \\ 
& = \frac12\left(C_{lq(1)} - C_{lq(3)} 
     - C_{qe} - C_{eu} + C_{lu}\right), 
\end{split} \\
\begin{split}
& C_{\text{PV}d} = C_{LVd}-C_{RVd} \\
& = \frac12\left(C_{lq(1)} + C_{lq(3)}
     - C_{qe} - C_{ed} + C_{ld}\right), 
\end{split}
\end{alignat}
\end{subequations}
as one can see by adding up $\Delta A_{\text{PV}}^{LVf}(\mathcal{N})$ and $\Delta A_{\text{PV}}^{RVf}(\mathcal{N})$ in Eqs.~\eqref{eq:delta-apv-EFT1}-\eqref{eq:delta-apv-EFT2} for up and down quarks, respectively.
From this we can group coefficients into smaller sets which have a non-distinguishable effect. Namely, 
\begin{itemize}
    \item $C_{lq(1)}$ and $C_{qe}$ have the same effect (equal contribution to up and down couplings) with opposite sign.
    \item $C_{eu}$ and $C_{lu}$ have the same effect (contribution only to up couplings) with opposite sign.
    \item $C_{ed}$ and $C_{ld}$ have the same effect (contribution only to down couplings) with opposite sign.
    \item $C_{lq(3)}$ forms its own group (contribution to up and down couplings with same magnitude but opposite sign).
\end{itemize}
Therefore, we plot only one example of each of these groups in comparison in \autoref{fig:smeftbounds}.
We find that the comparison of nuclei with protons is particularly good at resolving degeneracies between $C_{lq(3)}$ and other operators. However, the comparison between $p$ and $\carb$ can also be expected to distinguish well between up quark and down quark interactions, as seen in the plot of $C_{eu}$ against $C_{ed}$. Moreover, if the SM expectation marked by an asterisk turns out to be correct (see the discussion after Eq.\ (\ref{eq:disc})), we can see that the current best fit of APV, shown as dark green lines, can be expected to be ruled out by combining proton and $\carb$ measurements of P2.

\section{\label{sec:concl}Conclusion}
In our study of the potential of the P2 experiment to test parity-violating new physics induced by SMEFT operators or leptoquarks, we found that P2 can be expected to be competitive with existing collider searches for such new physics and in many cases to have a better sensitivity to  leptoquarks with masses above around \SI{2}{TeV}. For all single operator scenarios and most single leptoquark scenarios, bounds from APV experiments with $\caes$ can be exceeded by P2. We stress, however, that the complementarity of using different targets, for instance protons and $\carb$ at P2, as well as $\caes$ in APV will allow to better disentangle up and down quark interactions as illustrated in \autoref{fig:smeftbounds}. Moreover, a potential tension between theoretical and experimentally determined weak charges of $\caes$ could be either confirmed or resolved at 95\% CL by P2 data.

\section*{Acknowledgments}
We thank Sudip Jana for helpful discussions. 
I.B.\ acknowledges support by the IMPRS-PTFS. The work of B.D.\ is supported in part by the U.S.\ Department of Energy under Grant No.~DE-SC0017987. The work of Y.Z.\ is supported by the National Natural Science Foundation of China under Grant No.\  12175039, the 2021 Jiangsu Shuangchuang (Mass Innovation and Entrepreneurship)
Talent Program No.\ JSSCBS20210144, and the ``Fundamental Research Funds for the Central Universities''.  X.J.X is supported in part by the National Natural Science Foundation of China under Grant No. 12141501.


\appendix
\section{Mapping of SMEFT coefficients to the flavor basis}
\label{sec:app-flavorbasis}
If one starts with the SMEFT operators  $\ope_{lq(1)}$, $\ope_{lq(3)}$, $\ope_{lu}$, $\ope_{ld}$, $\ope_{qe}$, $\ope_{eu}$, $\ope_{ed}$, $\ope_{ledq}$, $\ope_{lequ(1)}$, and $\ope_{lequ(3)}$ in the flavor basis, after making the basis change
\be
\begin{split}
u_{L\alpha}' &= (V^{\dagger})_{\alpha\beta}\, u_{L\beta} \,, \\
u_{R\alpha}' &= u_{R\alpha}\,, 
\end{split}
\qquad
\begin{split}
d_{L\alpha}' &= d_{L\alpha}\vphantom{(V^{\dagger})_{\alpha\beta}} \,,\\
d_{R\alpha}' &= d_{R\alpha}\,, 
\end{split}
\qquad
\begin{split}
e_{L\alpha}' &= e_{L\alpha}\vphantom{(V^{\dagger})_{\alpha\beta}} \,,\\
e_{R\alpha}' &= e_{R\alpha}\, ,
\end{split}
\ee
with $V$ being the CKM matrix between primed flavor basis and unprimed mass basis, one finds
\be
\label{eq:formalism-charged-fermion}
\begin{split}
\lagr_{} &= \frac{1}{\Lambda^2}\left\{
(\widetilde{C}_{lq(1)}-\widetilde{C}_{lq(3)})(\overline{e}\gamma_\mu P_Le)(\overline{u}\gamma^\mu P_L u) \right. \\
&\quad +C_{lu}(\overline{e}\gamma_\mu P_Le)(\overline{u}\gamma^\mu P_Ru) \\
&\quad+(\overline{C}_{lq(1)}+\overline{C}_{lq(3)})(\overline{e}\gamma_\mu P_Le)(\overline{d}\gamma^\mu P_Ld) \\
&\quad +C_{ld}(\overline{e}\gamma_\mu P_Le)(\overline{d}\gamma^\mu P_Rd)
\\
&\quad 
+ C_{qe}(\overline{e}\gamma_\mu P_R e)(\overline{u}\gamma^\mu P_L u)
+ C_{eu}(\overline{e}\gamma_\mu P_R e)(\overline{u}\gamma^\mu P_R u)
\\
&\quad
+ C_{qe}(\overline{e}\gamma_\mu P_R e)(\overline{d}\gamma^\mu P_L d)
+ C_{ed}(\overline{e}\gamma_\mu P_R e)(\overline{d}\gamma^\mu P_R d)
\\
&\quad
-{C}_{lequ(1)}(\overline{e} P_Re)(\overline{u} P_R u)
-{C}_{lequ(1)}^{*}(\overline{e} P_L e)(\overline{u} P_L u)
\\
&\quad
+\, C_{ledq}(\overline{e} P_Re)(\overline{d} P_Ld)
+\, C_{ledq}^*(\overline{e} P_Le)(\overline{d} P_Rd)
\\
&\quad
-{C}_{lequ(3)}(\overline{e} \sigma_{\mu\nu} P_R e)(\overline{u} \sigma^{\mu\nu} P_R u) \\
&\quad \left. -{C}_{lequ(3)}^{*}(\overline{e} \sigma_{\mu\nu} P_L e)(\overline{u} \sigma^{\mu\nu} P_L u)
\right\}.
\end{split}
\ee
 The Wilson coefficients in Eq.~\eqref{eq:formalism-charged-fermion} are related to the Wilson coefficients $C_{j}^{\prime\abgd}$ defined for flavor eigenstates by the following identities 
for the parity-violating operators: 
\be
\begin{split}
\widetilde{C}_{lq(1)} &= V_{u\gamma}V^*_{u\delta}C_{lq(1)}^{\prime ee\gamma\delta}\,, \\
\overline{C}_{lq(1)} &= C_{lq(1)}^{\prime ee11}\,,\\ 
{C}_{lu} &= C_{lu}^{\prime ee11}\,, \\
{C}_{eu} &= C_{eu}^{\prime ee11}\,, \\
{C}_{qe} &= C_{qu}^{\prime ee11}\,. 
\end{split}
\qquad
\begin{split}
\widetilde{C}_{lq(3)} &= V_{u\gamma}V^*_{u\delta}C_{lq(3)}^{\prime ee\gamma\delta}\,, \\
\overline{C}_{lq(3)} &= C_{lq(3)}^{\prime ee11}\,, \\
{C}_{ld} &= C_{ld}^{\prime ee11}\,, \\
{C}_{ed} &= C_{ed}^{\prime ee11}\,, \\ 
\phantom{{C}_{qe}}
\end{split}
\ee
The notation of mass-basis coefficients is the same as it appears in Eq.~\eqref{eq:EFT-matching-nucleon}. 
Only two coefficients, $C_{lq(1)}$ and $C_{lq(3)}$ remain to be clarified. If we assume only first generation flavor couplings, we can set $\widetilde{C}_{lq(1)}=|V_{ud}|^2C_{lq(1)}^{\prime ee11}$ and $\widetilde{C}_{lq(3)}=|V_{ud}|^2C_{lq(3)}^{\prime ee11}$. Equating the Lagrangian of Eq.~\eqref{eq:formalism-charged-fermion} with the sum of Eq.~\eqref{eq:eff-lagrangian} plus Eq.~\eqref{eq:eff-lagrangian-spt}, we identify
\be
\begin{split}
C_{lq(1)}-C_{lq(3)} &= |V_{ud}|^2\left({C}_{lq(1)}^{\prime ee11}-\widetilde{C}_{lq(3)}^{\prime ee11}\right),\\
C_{lq(1)}+C_{lq(3)} &= \left({C}_{lq(1)}^{\prime ee11}-{C}_{lq(3)}^{\prime ee11}\right).
\end{split}
\ee
For the parity-conserving operators, we have
\be
\begin{split}
{C}_{lequ(1)} &= V^*_{u\gamma}C_{lequ(1)}^{\prime ee\gamma1}\,,\\
{C}_{lequ(3)} &= V^*_{u\gamma}C_{eluq(3)}^{\prime ee\gamma1}\,, \\
{C}_{ledq} &= C_{ledq}^{\prime ee11}\,.
\end{split}
\ee

\section{Parametrization of the form factors}
\label{sec:app-formfactors}

In this section we follow the notation of  Ref.~\cite{Becker:2018ggl}. The electromagnetic Sachs form factors of the proton can be parametrized by a model multiplying a dipole and a polynomial,
\begin{subequations}
\begin{alignat}{2}
G_E^p &= \left(1-\frac{q^2}{\SI{0.71}{GeV^2}}\right)^{-2}\left(1-\sum_{i=1}^8 \kappa_i^{E,p}q^{2i}\right) \,,\\
G_M^p &= \frac{\mu_\text{P}}{\mu_\text{N}}\left(1-\frac{q^2}{\SI{0.71}{GeV^2}}\right)^{-2}\left(1-\sum_{i=1}^8 \kappa_i^{M,p}q^{2i}\right)\,,
\end{alignat}
\end{subequations}
where $\mu_\text{P} = 2.792 847 356 \,\mu_\text{N}$ denotes the proton's magnetic
moment and $\mu_\text{N} = (e\hbar)/(2m_p)$ denotes the nuclear magneton. The fit coefficients $\kappa_i$ are given in Tables~17 and 18 of Ref.~\cite{Becker:2018ggl}. For the neutron form factors we use
\begin{subequations}
\begin{alignat}{2}
G_E^n &= \frac{\kappa_1^{E,n}\tau}{1+\kappa_2^{E,n}\tau}\left(1-\frac{q^2}{\SI{0.71}{GeV^2}}\right)^{-2}\,,\\
G_M^n &= \sum_{i=0}^9 \kappa_i^{M,n} q^{2i}\,,
\end{alignat}
\end{subequations}
where $\tau=-q^2/4m_p^2$ and 
the coefficients being given in Tables~19 and 20 of Ref.~\cite{Becker:2018ggl}. For the strangeness form factors we use 
\begin{subequations}
\begin{alignat}{2}
G_E^s &= \frac{\kappa_1^{E,s}\tau}{1+\kappa_2^{E,s}\tau}\left(1-\frac{q^2}{\SI{0.71}{GeV^2}}\right)^{-2}\,,\\
G_M^s &= \kappa^{M,s}_0 - \kappa_1^{M,s} q^2\,,
\end{alignat}
\end{subequations}
with the coefficients taken from Tables~21 and 22 of Ref.~\cite{Becker:2018ggl}.

\bibliographystyle{JHEP}
\bibliography{bibliography.bib}{}
\end{document}